\newcommand{\version}{September 3, 2013}
\documentclass[a4paper,twoside,11pt,pdftex]{article}
\usepackage[bindingoffset=0.4cm,textheight=22.5cm,hdivide={2.7cm,*,2.75cm}, vdivide={*,22cm,*}]{geometry}
\usepackage{amsmath,amsfonts,amssymb}
\usepackage[pdftex]{graphicx}
\usepackage{simplewick}			
\IfFileExists{dsfont.sty}
	{\usepackage{dsfont}
         \let\mathbb=\mathds
         \newcommand{\id}{\mathds{1}}}
	{\typeout{Package dsfont.sty was not found, using alternative macros.}
         \let\mathds=\mathbb
         \newcommand{\id}{\mbox{1 \kern-.59em {\rm l}}}}

\usepackage[pdftex,bookmarksnumbered=true,breaklinks=true]{hyperref}
\usepackage[numbers,square,comma,sort&compress]{natbib}

\newcommand{\uim}{UV/IR mixing}
\newcommand{\nc}{non-commu\-ta\-tive}

\newcommand{\etal}{\textit{et al.}}

\newcommand{\eqnref}[1]{Eqn.~(\ref{#1})}		
\newcommand{\figref}[1]{Fig.~\ref{#1}}			
\newcommand{\secref}[1]{Section~\ref{#1}}		
\newcommand{\appref}[1]{Appendix~\ref{#1}}		
\newcommand{\starco}[2]{\left[ #1\stackrel{\star}{,}#2\right] }		
\newcommand{\pa}{\partial}						
\newcommand{\diff}[2]{\frac{\pa #1}{\pa #2}}				
\newcommand{\ddiff}[3]{\frac{\partial^2 #1}{\partial #2 \partial #3}}   
\newcommand{\ri}{{\rm i}}						
\newcommand{\re}{{\rm e}}						
\newcommand{\p}{\tilde{p}}						

\newcommand{\g}{\gamma}
\renewcommand{\d}{\delta}

\newcommand{\vare}{\varepsilon}

\renewcommand{\th}{\theta}

\renewcommand{\l}{\lambda}
\newcommand{\m}{\mu}
\newcommand{\n}{\nu}

\newcommand{\ph}{\phi}

\newcommand{\G}{\Gamma}

  \newcommand{\cF}{\mathcal{F}}

  \newcommand{\cO}{\mathcal{O}}

\newcommand{\R}{\mathds{R}}

\newcommand{\inv}[1]{\frac{1}{#1}}				
\newcommand{\tinv}[1]{\tfrac{1}{#1}}
\newcommand{\intk}{\int\! d^4k}					
\newcommand{\intx}{\int\! d^4x}						
\newcommand{\nn}{\nonumber}

\newcommand{\wsq}{\widetilde{\square}}
\newcommand{\ig}{{\rm i}g}

\title{\begin{flushright}
        {\small LA-UR-13-24956}\vspace*{2em}
       \end{flushright}
BPHZ renormalization  and its application \\
to non-commutative field theory
}
\date{\version}

\author{Daniel N. Blaschke\footnotemark[1]~, Fran\c{c}ois Gieres\footnotemark[2]~, Franz Heindl\footnotemark[3]~,
\\Manfred Schweda\footnotemark[3]~ and Michael Wohlgenannt\footnotemark[4]}

\graphicspath{{./}{./figures/}}

\begin{document}
\maketitle
\thispagestyle{empty}
\begin{center}
\renewcommand{\thefootnote}{\fnsymbol{footnote}}
\vspace{-0.3cm}\footnotemark[1]Los Alamos National Laboratory, Theory Division\\Los Alamos, NM, 87545, USA\\[0.3cm]
\footnotemark[2]Universit\'e de Lyon, Universit\'e Lyon 1 and CNRS/IN2P3,\\Institut de Physique Nucl\'eaire, Bat. P. Dirac,\\4 rue Enrico Fermi, F-69622-Villeurbanne (France)\\[0.3cm]
\footnotemark[3]Institute for Theoretical Physics, Vienna University of Technology\\Wiedner Hauptstra\ss e 8-10, A-1040 Vienna (Austria)\\[0.5cm]
\footnotemark[4]Austro-Ukrainian Institute for Science and Technology,\\c/o TU Vienna, Wiedner Hauptstra\ss e 8-10, A-1040 Vienna (Austria)\\[0.5cm]
\ttfamily{E-mail: dblaschke@lanl.gov, gieres@ipnl.in2p3.fr, franz.m.heindl@gmail.com, mschweda@tph.tuwien.ac.at, michael.wohlgenannt@univie.ac.at}
\end{center}

\vspace{1.5em}
\begin{abstract}
\noindent
In a recent work a modified BPHZ scheme has been introduced and applied to one-loop Feynman graphs in {\nc} $\phi^4$-theory.
In the present paper, we first review the BPHZ method  and then we apply the modified BPHZ scheme as well as Zimmermann's forest formula to the sunrise graph, i.e. a typical higher-loop graph involving overlapping divergences.
Furthermore, we show that the application of the modified BPHZ scheme to the IR-singularities appearing in non-planar graphs (UV/IR mixing problem) leads to the introduction of a $1 /{p}^{\, 2}$ term and thereby to a renormalizable model.
Finally, we address the application of this approach to gauge field theories.
\end{abstract}


\newpage
\thispagestyle{empty}
\tableofcontents

%
\section{Introduction}
\label{sec:intro}
In this work we
continue the discussion of the BPHZ renormalization of
non-commutative $\phi^4$-theory in four Euclidean dimensions
initiated in reference~\cite{Blaschke:2012ex}
and we address in particular the issue of higher-loop graphs involving overlapping divergences.
The theory under consideration
is defined  at the classical level  by the  action
(e.g. see reference~\cite{Szabo:2001})
\begin{align}
\label{action}
\Gamma^{(0)}[\phi]=\Gamma^{(0)}_{\text{free}}[\phi]+\Gamma^{(0)}_{\text{int}}[\phi]
\equiv
\frac 12 \int d^4x \left( \partial^\mu \phi \star \partial_\mu \phi+m^2 \, \phi\star \phi
 \right)
 \, + \,
\frac \l{4!} \int d^4x \left(\phi\star\phi\star\phi\star\phi \right)\,.
\end{align}
From the definition of the Moyal star product,
\begin{align}
\left( f \star g \right) (x):= \left( \re ^{\frac{\ri}{2}\th^{\m\n}\pa^x_\m\pa^y_\n}f(x)g(y) \right) \Big|_{x=y}
\, , \qquad
{\rm with} \ \; \th^{\m\n} = - \th^{\n\m} \ \; {\rm constant} \, ,
\end{align}
it follows that
\begin{align}
\Gamma^{(0)}[\phi] = \int d^4x \left[ \frac 12 ( \partial^\mu \phi \, \partial_\mu \phi + m^2 \phi^2 )
+ \frac \l{4!} \, (\phi\star\phi\star\phi)(x) \, \phi(x) \right] \, .
\end{align}
Introducing the Fourier components
 $\tilde\phi(k)$  of $\phi$ by
$\phi(x)=\frac{1}{(2\pi)^4}\int d^4k \, \re ^{\ri kx}\tilde \phi(k)$,
we find that the propagator in momentum space is given by
\begin{align}
\label{propagator}
\tilde\Delta (k)=\frac{1}{k^2+m^2}
\, ,
\end{align}
and that the interaction term can be expressed
in terms of the variables $\tilde k_\mu \equiv \theta_{\mu\nu}k^\nu$ by
\begin{align}
\nonumber
\Gamma^{(0)}_{\text{int}}[\phi]=\frac{1}{4!} \intk_1\ldots d^4k_4 \, \tilde\phi(k_1)\tilde\phi(k_2)\tilde\phi(k_3)\tilde\phi(k_4) \, (2\pi)^4 \, \delta^{(4)}(k_1+k_2+k_3+k_4)
\, \bar{\lambda} \, ,
\end{align}
with
\begin{align}
\label{F9}
\bar\lambda \equiv \frac{\lambda}{3}\bigg[\cos\frac{k_1\tilde k_2}{2}\cos\frac{k_3\tilde k_4}{2}+\cos\frac{k_1\tilde k_3}{2}\cos\frac{k_2\tilde k_4}{2}+\cos\frac{k_1\tilde k_4}{2}\cos\frac{k_2\tilde k_3}{2}\bigg]
\, .
\end{align}
Henceforth, in comparison to the commutative $\phi^4$-theory, the interaction vertex of the {\nc} $\phi^4$-theory
is characterized by a modified coupling in momentum space
($\lambda$ becomes $\bar{\lambda} $).

The quantization of this model and the renormalization of related scalar field models
has been discussed over the last fifteen years, see for instance reference~\cite{Blaschke:2012ex}
for a brief review and list of references.
In the latter work it was pointed out that the usual BPHZ momentum space subtraction scheme (which consists
of subtracting
appropriate polynomials in the external momentum from the integrand of
divergent integrals)
cannot be applied in non-commutative  theories, e.g. for an integral of the form
\begin{align}
\label{4pfct}
J (p) \equiv
 \intk \,  \frac{\cos(k\p)}{[(p+k)^2+m^2][k^2+m^2]}
 \,.
\end{align}
The problem is due to the phase factor $\cos(k\p)$ which is at the origin of the UV/IR mixing problem,
i.e. the appearance of an IR-singularity for small values of the external momentum $p$.
Therefore, a modified subtraction scheme was proposed in reference~\cite{Blaschke:2012ex}:
it consists of considering $p$ and $\p$ as independent variables (though satisfying $p \tilde p =0$) when performing the subtraction,
i.e. one subtracts from the integrand
its Taylor series expansion with respect to the external momentum $p$ around $p=0$ up to the order
of divergence of the graph,
while maintaining the phase factors: thus, for the integral (\ref{4pfct})  one considers
\begin{align}
\label{checkJ}
J^{{\rm finite}} (p)
\equiv
\intk\left( \frac{\cos(k\p)}{[(p+k)^2+m^2][k^2+m^2]} - \frac{\cos(k\p)}{[k^2+m^2]^2} \right)
 \,.
\end{align}
By proceeding in this way, the one-loop renormalization of the theory could be carried out for the so-called na\"ive $\phi^4$-theory
described by the action (\ref{action})
as well as for this action supplemented by a $1/p^2$-term which is known to overcome the UV/IR mixing problem
while maintaining the translation invariance of the model~\cite{Gurau:2009}.

Part of the present work
concerns the application of the modified subtraction scheme to
 higher-loop graphs involving overlapping divergences.
For concreteness, we will investigate in detail
the \emph{sunrise graph} as an example for a two-loop graph with overlapping divergences.
To tackle this problem, we apply the so-called forest formula of Zimmermann~\cite{Zimmermann:1969}
in the {\nc} setting using the modified subtraction scheme.

In this context, it is worthwhile recalling that an open problem in  {\nc} field theories
is that there exists no proof for the renormalizability of the proposed classical gauge field models.
Indeed, the methods considered so far for the quantization of scalar field theories on {\nc} space
(such as multi-scale analysis) cannot be applied, or at least not without some serious complications,
to gauge field theories due to the fact that they break the gauge symmetry. This is a strong impetus
 for trying to generalize to the {\nc} setting the approach
of BPHZ which does not require the introduction of a regularization
and which has been proven to be a powerful tool for field theories with local symmetries
on commutative space, e.g. see references~\cite{Piguet:1995,Schweda-book:1998}.
With this motivation in mind and in order to present a self-contained treatment
of the sunrise-graph, we provide a short introduction to the BPHZ-approach to renormalization
in \secref{sec:nutshell} before treating the sunrise-graph of {\nc} $\ph^4$-theory in \secref{sec:ffphase} and discussing IR-singularities in \secref{IRsingularities}.
We conclude with some comments on gauge theories and the issues to be addressed in future work.

\section{BPHZ in a nutshell}\label{sec:nutshell}

In this section we present a sketch of the \emph{BPHZ (Bogoliubov, Parasiuk, Hepp, Zimmermann) method
of renormalization} in Euclidean $\R ^4$
which is based on some ideas put forward by Stueckelberg and Green~\cite{stueckelgreen,Wightman:1975}.
The original references
are~\cite{Bogoliubov:1955,Bogoliubov:1956,Bogoliubov:1957gp,Bogoliubov:1980,Hepp:1966eg, Zimmermann:1967}\cite{Zimmermann:1969}\cite{Lowenstein:1975}
and the approach has been conveniently reformulated by
Zimmermann~\cite{Zimmermann:1967,Zimmermann:1969, ZimmermannBrandeis},
different aspects being discussed in
references~\cite{ZimmermannAnnals,Lowenstein:1974uk, Lowenstein:1974qt, Lowenstein:1975ug, Lowenstein:1975pd,
Schweda:1982ny,Itzykson:2005,Manoukian:1983,Collins:1986,Zavyalov:1990,Smirnov:1991,muta,das,CollinsArticle,sibold,fredenhagen}.

\subsection{Generalities}

The BPHZ approach to the perturbative renormalization of field theory
amounts to a proper definition of the quantum theory and
essentially consists of three ingredients:
\begin{enumerate}
\item A \emph{subtraction scheme}, i.e
a systematic
procedure for subtracting an overall ultraviolet divergence as well as subdivergences (i.e. divergences of subdiagrams)
from any Feynman integral
(1PI (one-particle-irreducible) graph or
 function) in order to get a finite (convergent) expression for this integral.
\item A \emph{proof of the locality of these subtractions}, i.e.
a proof that these subtractions correspond to the addition of local counterterms
to the Lagrangian having the same form as those in the original Lagrangian\footnote{
In the {\nc} 
setting to be considered in Sections~\ref{nctheory} and~\ref{ncgt}, 
this condition obviously has to be relaxed since the star product introduces 
a certain type of non-localities, see Ref.~\cite{Blaschke:2012ex}.
However, the condition that the counterterms have 
the same form as those in the original Lagrangian still applies -- see also the discussion in \secref{sec:modifiedBPHZ}.}
(compatibility with additive renormalization).
\item
A set of \emph{normalization conditions} on the 1PI
functions involving divergences in order to fix the physical parameters
(masses, coupling constants, \ldots) of the theory.
These conditions can be achieved by the addition of finite counterterms to the
Lagrangian.
\end{enumerate}
If the initial Lagrangian has some rigid or local \emph{symmetries},
these also have to be taken into account, i.e. the counterterms have to be invariant
(and if a regularization is chosen for performing the subtractions, it
has to respect the symmetries).
A systematic method for implementing this program is provided by the
method of \emph{algebraic renormalization}~\cite{Piguet:1995,Schweda-book:1998}.

In the BPHZ  approach, the subtraction of divergences can be performed by considering a particular
regularization (such as dimensional regularization)
or without using such a regularization (momentum space subtraction in the integrands of Feynman integrals).
The latter approach allows to treat in particular
field theories for which no invariant regulator is known.

In the BPHZ method, the systematic subtraction of subdivergences for a given graph is realized by Zimmermann's \emph{forest formula} which
yields finite (i.e. renormalized) Feynman integrals. The fact that the latter formula
represents a solution of the \emph{recursion relation} for Bogoliubov's \emph{$R$-operation}
allows to prove inductively the locality of the subtractions.
 Thus, the BPHZ approach to renormalization is equivalent to the familiar method based on
counterterms.  Different renormalization schemes are always
related by a finite renormalization.

\subsection{Subtraction operator}\label{subtractoperator}

\begin{figure}[ht]
\centering
\includegraphics[width=0.4\columnwidth]{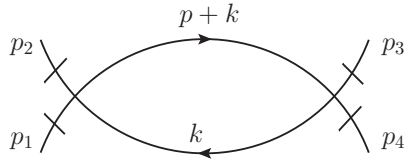}
\caption{The fish diagram}
\label{fig:fishdiagram}
\end{figure}

Suppose we have an  integral $J_{\Gamma}$ corresponding to
an amputated 1PI Feynman diagram $\Gamma$
which has an overall divergence, but no subdivergence.
(The treatment of
graphs containing divergent subgraphs
will be addressed in Subsection~\ref{sec:treatsubdiv}.)
An example~\cite{Piguet:1995}
concerning the $\phi^4$-theory
in Euclidean $\R ^4$ is given by the one-loop four-point 1PI graph
(the so-called \emph{fish diagram} depicted in \figref{fig:fishdiagram})
depending on the total incoming external
momentum $p =p_1 +p_2 = p_3 +p_4$:
\begin{align}
J_{\Gamma} (p) \equiv
 \intk \,  \tilde{\Delta} (p+k) \,  \tilde{\Delta} (-k) =
 \intk \,  \frac{1}{[(p+k)^2+m^2][k^2+m^2]}
\equiv  \intk \, I_{\Gamma} (p,k)
 \,.
 \label{4ptfct}
\end{align}
This integral is logarithmically divergent, i.e.  the \emph{superficial degree of divergence} $\delta (\Gamma )$
of the graph $\Gamma$ vanishes.
A finite part for such an integral satisfying
$\delta (\Gamma ) \geq 0$ can be extracted in different manners.
Quite generally, for an integral
$J_{\Gamma} (\underline{p} ) \equiv \intk \, I_{\Gamma} (\underline{p} ,k) $
depending on the external momenta $\underline{p} \equiv (p_1, \dots, p_n) $
with $p_1 + \dots + p_n =0$,
the \emph{BPHZ momentum space subtraction}
is defined by\footnote{More generally, one may have integrals over several internal momenta
$k_1, \dots , k_L$.}
\begin{align}
\label{standardBPHZ}
\fbox{\mbox{$ \
J_{\Gamma}^{{\rm finite}} (\underline{p})
\equiv  \intk \, \left[ 1 - t_{\underline{p}}^{\delta(\Gamma) } \right] \,  I_{\Gamma} (\underline{p},k)
\ $}}
\, ,
\end{align}
where the \emph{Taylor expansion operator} (acting on the integrand of the graph $\Gamma$) is given by
\begin{align}
\label{teo}
\left( t_{\underline{p}} ^{N} I_{\Gamma} \right) (\underline{p} ,k)
\equiv \sum_{l=0}^{N} \frac{1}{l!} \, p_{i_1}^{\mu_1} \cdots p_{i_l}^{\mu_l}
\, \frac{\partial^l I_{\Gamma}}{\partial p_{i_1}^{\mu _1} \cdots \partial p_{i_l}^{\mu _l} }\big( \underline{p} =\underline{0} ,k \big)
\, .
\end{align}
In particular, for a single external momentum, i.e. $ \underline{p} = p$, one has
\begin{align}
\left( t_{p} ^{N} I_{\Gamma} \right) ( p ,k)
= I_{\Gamma} (0,k) +  p^{\mu}
\, \frac{\partial I_{\Gamma}}{\partial p^{\mu } } \big( 0 ,k \big)
+ \dots + \frac{1}{N!} \, p^{\mu_1} \cdots p^{\mu_N}
\, \frac{\partial^N I_{\Gamma}}{\partial p^{\mu _1} \cdots \partial p^{\mu _N} } \big( 0 ,k \big)
\, .
\label{eq:subtact-op-onep}
\end{align}
Thus, for the integral (\ref{4ptfct}), we obtain
\begin{align}
\label{sublog}
J_{\Gamma}^{{\rm finite}} (p)
 \equiv  \intk \, [1 - t_{p}^{0}  ] \,  I_{\Gamma} (p,k)
 & = \intk \, [ I_{\Gamma} (p,k) -  I_{\Gamma} (0,k) ]
 \\
 &= \intk \, \left(  \frac{1}{[(p+k)^2+m^2][k^2+m^2]} -  \frac{1}{[k^2+m^2]^2}  \right)
 \,,
 \nn
\end{align}
which represents a convergent integral.
In this respect, we note that
$I_{\Gamma} (p,k) - I_{\Gamma} (0,k) = p^{\mu}
\, \frac{\partial I_{\Gamma}}{\partial p^{\mu} } \big( 0 ,k \big) + \cdots$
and that the \emph{differentiation with respect to $p^{\mu}$  lowers the degree
of divergence of the propagator} and thereby of the integral.
For the logarithmically divergent integral \eqref{4ptfct},
the first derivative already renders the integral UV-convergent:
\begin{align}
 \frac{\partial J_{\Gamma}}{\partial p^{\mu} } \big( p \big) =
 - \intk \, \frac{2(p+k)_{\mu}}{[(p+k)^2+m^2]^2 \, [k^2+m^2]}
 \sim \int \frac{dK \, K^4 }{K^6} \sim K^{-1} \, .
 \end{align}
In fact~\cite{Piguet:1995}, the latter expression may be taken as the \emph{definition of the renormalized
integral associated to} $J_{\Gamma}(p)$; since it follows from (\ref{sublog}) that
$\frac{\pa J_{\Gamma}^{{\rm finite}}}{\pa p^{\mu}} (p)
= \frac{\pa J_{\Gamma}}{\pa p^{\mu}} (p)$, we conclude that $J_{\Gamma}^{{\rm finite}}$ is only defined
up to a real constant $d$. (Equivalently, we can argue that the Taylor series expansion of $[1 - t_{p}^{0}  ] \,  I_{\Gamma} (p,k)$
with respect to $p^{\mu}$ vanishes to order zero.)
Henceforth, we have the freedom of considering
\begin{align}
\label{addconst}
J_{\Gamma}^{{\rm finite}} (p) \leadsto J_{\Gamma}^{{\rm finite}} (p) +d
\, , \qquad (d \in \R )
\, .
\end{align}
From the fact that $J_{\Gamma} (p)$ represents the one-loop four-point function,
one concludes that the addition (\ref{addconst}) in momentum space amounts to adding to the Lagrangian
a \emph{finite counterterm} having the same form as the original interaction term:
\begin{align}
{\cal L} \leadsto {\cal L} +d \, \phi^4
\, .
\end{align}
This counterterm originating from the one-loop subtraction, is of order $\hbar$.

Summarizing the considered example and coming back to the general case of an integral
$J_{\Gamma} (p) \equiv \intk \, I_{\Gamma} (p,k)$ with superficial degree of divergence
$\delta (\Gamma )$, we can say the following.
The operation $t_p ^{\delta(\Gamma)}$ allows to extract  from the integrand
$I_{\Gamma}$ the divergent part of the integral $J_{\Gamma}$:
this part is a polynomial in the external momenta whose order coincides
with the degree of divergence $\delta (\Gamma )$ of the graph $\Gamma$.
The operator
\begin{align}
R_{\Gamma} \equiv 1 - t_p ^{\delta(\Gamma)}
\label{R}
\end{align}
which extracts a finite part from $J_{\Gamma}$ is   Bogoliubov's \emph{$R$-operation}
for the case of an integral having only an overall divergence (which we consider in this subsection).
The renormalized integral associated to $J_{\Gamma}$
can be defined as its momentum space derivative of order $\delta (\Gamma ) +1$ which represents a convergent integral.
The extraction of a finite part
from $J_{\Gamma} (p)$
by means of  Bogoliubov's $R$-operation yields the expression
$J_{\Gamma}^{{\rm finite}} (p)$
which is only determined up to a polynomial in $p$ of degree $\delta (\Gamma )$:
\begin{align}
\label{addpolynom}
J_{\Gamma}^{{\rm finite}} (p) \leadsto J_{\Gamma}^{{\rm finite}} (p) +
\sum_{l=0}^{\delta (\Gamma)} \frac{1}{l!} \, d_{\mu_1 \dots \mu_l}\, p^{\mu_1} \cdots p^{\mu_l}
\, .
\end{align}
In configuration space the latter polynomial in $p$ corresponds
to \emph{finite local counterterms} involving $\pa^{\mu_1} \phi, \dots , \pa^{\mu_l} \phi$.

It is instructive~\cite{Piguet:1995} to investigate the one-loop two-point 1PI graph
of the $\phi ^4$-theory (the so-called \emph{tadpole} graph)
from this viewpoint since it does not depend on the external momentum
and is quadratically divergent (i.e. $\delta (\Gamma ) =2$): up to a constant factor, it is given by
\begin{align}
J_{\Gamma} = \int \frac{d^4 k}{k^2 +m^2} \equiv \intk \, I_{\Gamma} (p,k)
\, .
\end{align}
Application of the BPHZ subtraction scheme yields a vanishing result:
\begin{equation}
\label{finitenpltadpole}
J_{\Gamma}^{{\rm finite}}
 \equiv  \intk \, [1 - t_{p}^{2}  ] \,  I_{\Gamma} (p,k)
  = \intk \, [ I_{\Gamma} (p,k) -  I_{\Gamma} (0,k) ]
 =0
 \, .
\end{equation}
However, as argued above, this expression is determined up to a quadratic polynomial in $p$.
Taking into account Lorentz invariance, the latter polynomial reduces to $a + c p^2$
and amounts to adding to the Lagrangian finite counterterms $a \phi^2 + c (\pa^{\mu}\phi)(\pa_{\mu}\phi)$
i.e. terms having the same form as the original kinetic and mass terms.
The finite constants $a,\, c, \, d$ can be used to adjust the two-point and four-point functions to given normalization
conditions.

While the described subtraction procedure has the advantage of not using a particular regularization,
for the $\phi^4$-theory
one may as well consider such a regularization, e.g. dimensional regularization involving the complex parameter
$\varepsilon \equiv \frac{4-D}{2}$, where $D$ denotes the space(-time) dimension (which eventually approaches 4).
A regularized divergent integral is then given by a Laurent series:
$J_{\Gamma} (\varepsilon) = \sum_{n=-N} ^{\infty} a_n \varepsilon^n$ with $N$ being the number of loops in $\Gamma$.
The renormalization prescription now consists of subtracting  the negative powers  in the Laurent series
and subsequently removing the regulator, i.e. the formula $J_{\Gamma}^{{\rm finite}} = R_{\Gamma} J_{\Gamma}$
means
\begin{align}
J_{\Gamma} ^{{\rm finite}}
 &= \lim_{\varepsilon \to 0} \left[ 1-  t^{\delta(\Gamma) } \right]   J_{\Gamma} (\varepsilon)
 \equiv \lim_{\varepsilon \to 0} \left[  J_{\Gamma} (\varepsilon) -  J_{\Gamma}^{{\rm divergent}} (\varepsilon) \right]
\nn
  \\
  &= \lim_{\varepsilon \to 0} \, \sum_{n=0 } ^{\infty} a_n \varepsilon^n = a_0
  \, .
 \end{align}
The ambiguity in extracting a finite part by different methods corresponds to a finite renormalization.

Before dealing with the systematic subtraction of divergences of subgraphs, we look at the zoology of the latter
graphs~\cite{ZimmermannBrandeis}.

\subsection{Subgraphs and forests}\label{sec:subgraphs}
\begin{figure}[ht]
 \centering
 \includegraphics[width=0.91\columnwidth]{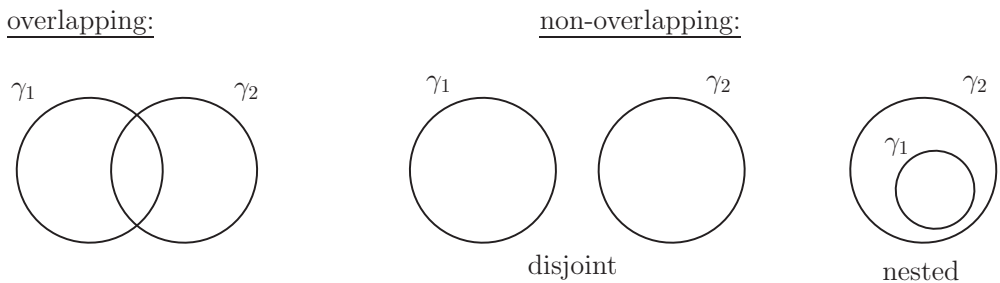}
 \caption{Types of subgraphs}
\label{fig:nonoverlapping}
\end{figure}

For a given graph $\Gamma$, a $1PI$
subgraph $\gamma\subset\Gamma$
may be superficially divergent, i.e. its superficial  degree of divergence $ \delta (\gamma)$ satisfies
$\delta (\gamma)\geq 0$: this subgraph is then called a \emph{renormalization part} of $\Gamma$.

For a given graph $\Gamma$, we may have different types of subgraphs $\gamma$:
There may be overlapping graphs and non-overlapping ones, and among the latter we can distinguish the disjunct and the nested ones.
This situation is best summarized~\cite{ZimmermannBrandeis} by \figref{fig:nonoverlapping}.

More precisely, consider two subgraphs $\gamma_1$ and $\gamma_2$.
If $\gamma_1$ is completely included in $\gamma_2$ as a subgraph ($\gamma_1 \subset \gamma_2$),
then $\gamma_1$ is said to be \emph{nested in $\gamma_2$} and $\gamma_2$ is called a \emph{nested
graph}.
The subgraphs $\gamma_1$ and $\gamma_2$ are said to be \emph{disjunct}  $( \gamma_1\cap\gamma_2= \emptyset )$ if they have no line or vertex in common.
The subgraphs $\gamma_1$ and $\gamma_2$ are referred to as \emph{non-overlapping} if one  of the following conditions is satisfied:
\begin{align}
\gamma_1&\subset \gamma_2 \, , \quad
\gamma_2 \subset \gamma_1 \, , \quad
\gamma_1 \cap\gamma_2= \emptyset
\, .
\end{align}
Otherwise these subgraphs are called \emph{overlapping}.
In the latter case, the diagrams have some common internal lines and vertices, and their union is referred to as
the \emph{overlapping diagram}; the divergence resulting from such a diagram is called the \emph{overlapping divergence}.
Due to their very nature, the elimination of  such divergences requires two different subtractions.
This fact represents a serious problem (the infamous
``overlap problem'') for a recursive proof of the renormalizability for a given theory if one uses a method
(like Dyson's original counterterm method) in which all subgraphs, including the overlapping ones,
have to be taken into account. As we will see in \secref{sec:treatsubdiv}, the BPHZ method (and more precisely the forest formula)
circumvents these problems since it only requires to deal with non-overlapping subgraphs.

For a graph $\Gamma$, one can introduce different sets of subgraphs $\gamma \subset \Gamma$
where $\gamma$ may be the full graph $\Gamma$ or the empty graph $\emptyset$.  These sets are generically referred to as \emph{forests}
and in our overview we do not discuss the classification of forests.

The unrenormalized integrand
corresponding to the graph $\Gamma$ can be decomposed with respect to the one of a subgraph $\gamma \subset \Gamma$ according to
\begin{align}
I_\Gamma(p,k) =I_{\Gamma/\gamma}I_\gamma(p^\gamma,k^\gamma)
\, .
\end{align}
Here the so-called \emph{reduced diagram} $\Gamma/ \gamma $ is obtained from $\Gamma$ by contracting
$\gamma$ to a point. Moreover,
the internal and external momenta $k^\gamma$ and $p^\gamma$ of the subgraph $\gamma$ have to be chosen
to be consistent with the ones parameterizing the graph $\Gamma$
and with the energy-momentum conservation at the external vertices of the subgraph.
This choice can be formalized by the introduction of the so-called \emph{substitution operator} $S_\gamma $ \cite{Zimmermann:1969}
whose action is best illustrated by considering the example of the subgraph $\gamma_1$ of the sunrise graph $\Gamma$, as we
will do in the following subsection.

\subsection{Example: subgraphs of the sunrise graph}\label{sec:exsun}

\begin{figure}[ht]
\centering
\includegraphics[width=\columnwidth]{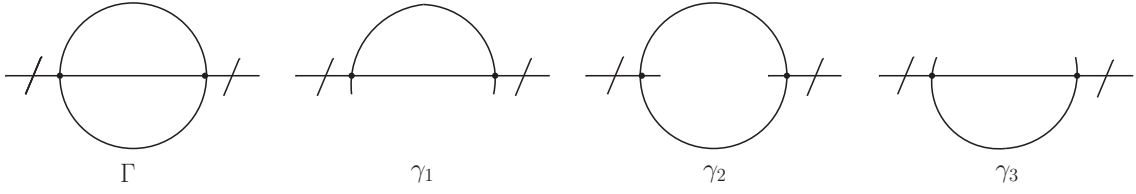}
\caption{Sunrise graph $\Gamma$ and subgraphs $\gamma_i$}
\label{fig:Sunrise-graph-and-subgraphs}
\end{figure}

As an example of a graph with overlapping divergences we consider the case of the
sunrise graph in $\phi ^4$-theory which is
depicted in \figref{fig:Sunrise-graph-and-subgraphs}.
$\Gamma$ represents the full sunrise graph with $\delta (\Gamma)=2$, whereas the $\gamma_i$ are the
non-trivial subgraphs all of which satisfy $\delta (\gamma_i)=0$.

\begin{figure}[ht]
\centering
\includegraphics[width=0.45\columnwidth]{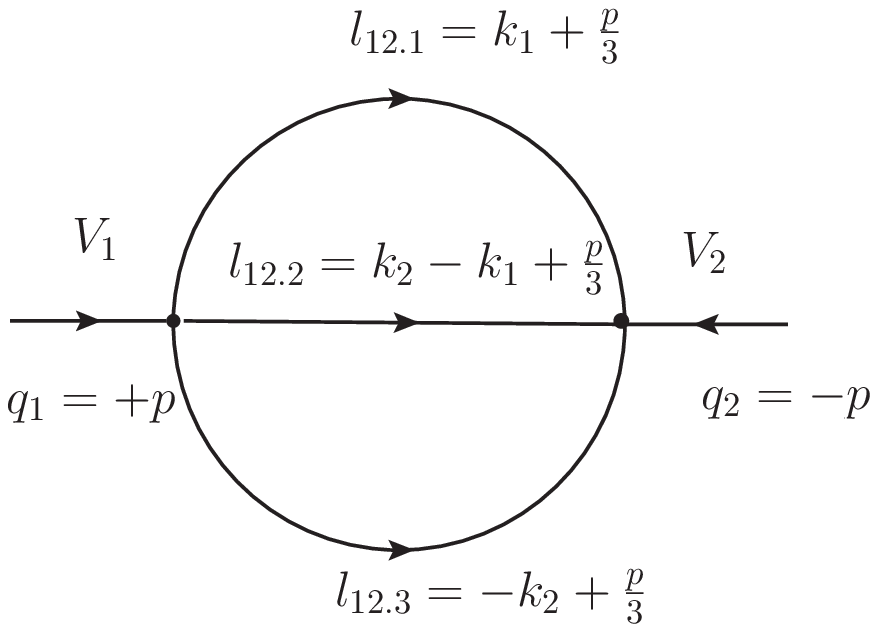}
\quad
\includegraphics[width=0.45\columnwidth]{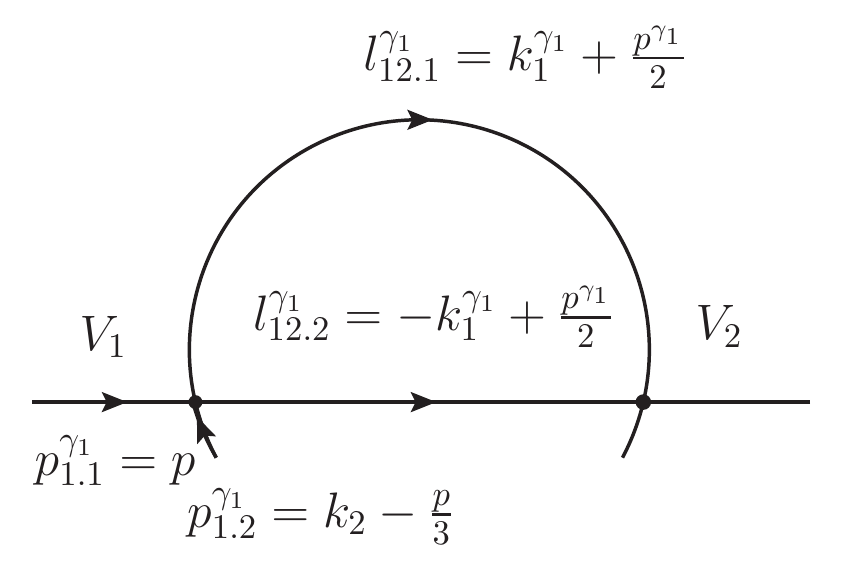}
\caption{Sunrise graph $\Gamma$ and subgraph $\gamma_1$ with momenta}
\label{fig:Sunrise-with-momenta}
\end{figure}

Let us make~\cite{ZimmermannBrandeis,Schweda:1982ny}
the assignment of momenta for the sunrise graph shown in \figref{fig:Sunrise-with-momenta}.
Here, $l_{12.i}$ with $i \in \{ 1,2,3\}$ denotes the different momenta flowing from vertex $V_1$ to vertex $V_2$.
For the subgraph $\gamma_1$ one chooses the assignment of momenta
shown in \figref{fig:Sunrise-with-momenta} (the \emph{external} momenta $p_{1.1}^{\gamma_1}$ and $p_{1.2}^{\gamma_1}$
at the vertex $V_1$
of the subgraph $\gamma_1$ being determined by the corresponding momenta of the graph $\Gamma$).
The values of the internal momenta of the subgraph  $\gamma_1$
are determined by  looking at vertex $V_1$ and
by comparing with the momenta in graph $\Gamma$:
\begin{align}\label{F21}
&p^{\gamma_1}     \equiv     p_{1.1}^{\gamma_1}+p_{1.2}^{\gamma_1}
\stackrel{!}{=} p+ \left( k_2-\frac{p}{3} \right) =k_2+\frac{2p}{3}
\,, \nn\\
&k_1^{\gamma_1}+\frac{p^{\gamma_1}}{2} \stackrel{!}{=} k_1+\frac{p}{3}
\, .
\end{align}
By combining these two relations we get
\begin{align}
k_1^{\gamma_1}+\frac{1}{2}\left( k_2+\frac{2p}{3} \right) \stackrel{!}{=} k_1+\frac{p}{3}
\, ,
\end{align}
i.e. in summary
\begin{align}
\label{kgamma1}
p^{\gamma_1}&=k_2+\frac{2p}{3}\,, \qquad
k_1^{\gamma_1} =k_1-\frac{k_2}{2}
\,.
\end{align}
We note that this assignment of momenta is consistent with the one of the ``horizontal'' line
of the graphs $\Gamma$ and $\gamma_1$:
\begin{align}
\frac{p^{\gamma_1}}{2}-k^{\gamma_1}_1=\frac{k_2}{2}+\frac{p}{3}-k_1+\frac{k_2}{2}=k_2-k_1+\frac{p}{3}.
\end{align}
This illustrates how the assignment of momenta (or equivalently the action of the substitution operator) works.
We note that
the reduced diagram ${\Gamma /\gamma_1}$  is the tadpole graph with loop momentum $-k_2 +\frac{p}{3}$
so that the corresponding
integrand is given  by
$I_ {\Gamma /\gamma_1}
= \tilde{\Delta} (-k_2 +\frac{p}{3})$
where $\tilde{\Delta}$ denotes the momentum space propagator (\ref{propagator}).
For later reference, we spell out the
analog of $k_1^{\gamma_1}$ in (\ref{kgamma1}) for the subgraphs $\gamma_2$ and $\gamma_3$, respectively:
\begin{align}
\label{g2g3}
k_2^{\gamma_2} = \inv2 (k_1 + k_2 ) \, , \qquad k_3^{\gamma_3} = -k_2 + \inv2 \, k_1
\, .
\end{align}

\subsection{Subtraction of subdivergences: Forest formula}\label{sec:treatsubdiv}

Suppose the Feynman graph $\Gamma$ contains one or several renormalization parts $\gamma$ with $\gamma \neq \Gamma$.
To simplify the notation, the Taylor expansion operator for the graph $\gamma$,
which is defined as in equation (\ref{teo}) and which picks out a potentially divergent part of $\gamma$, is written
for short as $t_{\gamma}$.

A finite Feynman integral can be obtained for the graph $\Gamma$
by subtracting all divergences of the integral from its integrand $I_{\Gamma}$:
this is achieved by  applying the subtraction operator $1 - t_{\gamma}$
for all  renormalization
parts $\gamma$, i.e. Bogoliubov's $R$-operation \eqref{R} now acts on $I_{\Gamma}$ according to\footnote{Actually this formula is tantamount to Dyson's original prescription of renormalization for non-overlapping divergences~\cite{Dyson:1949ha}.}
\begin{align}
\fbox{\mbox{$ \
R_\Gamma I_\Gamma
\equiv (1 - t_{\Gamma} ) \Big( \prod _{\gamma \in {\cal F}} (1 - t_{\gamma} ) \Big) I_{\Gamma}
\ $}}
\, .
\label{RR}
\end{align}
Here, the subtraction $1-t_{\Gamma}$ allows to eliminate an overall divergence
from the graph $\Gamma$: if none is present, the action of $t_{\Gamma}$ gives zero.
The set (``forest'') $\cF$ consists of all renormalization parts $\gamma$ with $\gamma \neq \Gamma$
and it is understood that \emph{the subtractions are performed from inside out,}
i.e. for nested subgraphs $\gamma \subset \gamma^{\prime}$,
the subtraction $1 - t_{\gamma}$ is performed before $1 - t_{{\gamma}^{\prime}}$.
The procedure (\ref{RR})
is an algorithmic method which  can always be applied to
eliminate potentially divergent contributions from a given graph and to obtain thereby
 absolutely convergent Feynman integrals (Weinberg's power counting theorem).

An important result established on general grounds by Berg\`ere and Zuber~\cite{Bergere:1974zh} is that overlapping subgraphs
can be discarded from formula (\ref{RR}).
More precisely, suppose $\gamma_1$ and $\gamma_2$ are overlapping renormalization parts of $\Gamma$; then, one has
\begin{align}
(1 - t_{{\gamma}_{12}} ) t_{\gamma_1} t_{\gamma_2} =0
\, ,
\label{cancel}
\end{align}
where ${\gamma}_{12}$ is a suitable renormalization part of $\Gamma$ which contains both
$\gamma_1$ and $\gamma_2$ as subgraphs. More generally, relation (\ref{cancel}) holds if the product $t_{\gamma_1} t_{\gamma_2}$
involves more factors corresponding to overlapping renormalization parts.
In the approach to renormalization based on counterterms, the result (\ref{cancel}) means that
the overlapping divergences (which can lead to non-local counterterms) are always eliminated by lower order counterterms
(see references~\cite{muta,das} for some explicit calculations) and therefore do not require the introduction of specific counterterms
(which would correspond to a non-vanishing subtraction $(1 - t_{{\gamma}_{12}} ) t_{\gamma_1} t_{\gamma_2}$).

As a simple \textbf{example}, assume that $\Gamma = \gamma_1 \cup \gamma_2$ where $\gamma_1$ and $\gamma_2$ are overlapping subgraphs
of the divergent graph $\Gamma$.  Then, relation (\ref{cancel})
yields $ ( 1 - t_{\Gamma} ) t_{{\gamma}_{1}}  t_{{\gamma}_{2}} =0$ so that expression
(\ref{RR}) can be rewritten as
\begin{align}
\label{exoverlap}
R_{\Gamma}  I_{\Gamma}  &\equiv  ( 1 - t_{\Gamma} ) (1 - t_{\gamma _1})(1- t_{\gamma _2} )  I_\Gamma
 = ( 1 - t_{\Gamma} ) (1 - t_{\gamma _1} - t_{\gamma _2} )  I_\Gamma
\, .
\end{align}

Quite generally, if we substitute equation (\ref{cancel}) and its generalizations involving more factors $t_{\gamma_i}$
 (corresponding to overlapping subgraphs $\gamma_i$)  into relation (\ref{RR}), we conclude that
\begin{align}
\label{forestformula}
\fbox{\mbox{$ \
R_\Gamma  I_\Gamma   = ( 1 - t_{\Gamma} ) \sum_{\alpha} \Big( \prod _{\gamma \in {\cal F}_{\alpha}} (- t_{\gamma} )  \Big) I_\Gamma
\ $}}
\, .
\end{align}
Here, the index $\alpha \in \{ 0,1, \dots \}$ labels all sets (forests) ${\cal F}_{\alpha}$ of
renormalization parts $\gamma \subset \Gamma, \, \gamma \neq \Gamma$,
 each of these sets containing only \emph{non-overlapping} subgraphs.
Moreover, one also takes into account the empty
forest ${\cal F}_0 \equiv \{ \emptyset \}$ given by the empty set for which one sets
$-t_{\emptyset} \equiv 1$.
Relation (\ref{forestformula}) is referred to as Zimmermann's \emph{forest formula}\footnote{The forest formula
appears to have been  discovered independently by Zavyalov and Stepanov~\cite{zavstep}, the BPHZ method
having been elegantly reformulated by Zimmermann~\cite{CollinsArticle,muta}.}  (for Bogoliubov's $R$-operation).

In example (\ref{exoverlap}) one has
 ${\cal F}_0 = \{ \emptyset \}, \, {\cal F}_1 = \{ \gamma_1 \}, \, {\cal F}_2 = \{ \gamma_2 \}$.
 Similarly for the sunrise graph in $\phi^4$-theory, we have
 ${\cal F}_0 = \{ \emptyset \}$ and  ${\cal F}_i = \{ \gamma_i \}$ for $i = 1,2,3$ so that
 the forest formula for the sunrise graph reads
\begin{align}
R_\Gamma I_{\Gamma}
 = ( 1 - t_{\Gamma} ) \;
\Big( 1 + \sum_{i=1}^3 (- t_{\gamma_i} ) \Big) I_{\Gamma}
\, .
\end{align}
The  subtraction operator $t_{\gamma_i}$ only affects the subgraph $\gamma_i$ of $\Gamma$
and $t_{\gamma_i}  I_{\gamma_i}$ amounts to picking out the divergent part of $\gamma_i$ (i.e.
generating the local counterterm which allows to subtract the divergence due to $\gamma_i$).
Thus, the effect of $t_{\gamma_i}$ on $ I_\Gamma$ is to contract the subgraph $\gamma_i$ to a point (thereby yielding the
reduced graph $\Gamma / \gamma_i$)
and to pick out the divergent part of $\gamma_i$:
\begin{align}
\label{fac}
 t_{\gamma_i}   I_\Gamma = I_{\Gamma / \gamma_i} \; t_{\gamma_i}  I_{\gamma_i}
 \, .
\end{align}
Henceforth, the \emph{forest formula for the sunrise graph} reads~\cite{Schweda:1982ny}
\begin{align}
R_\Gamma I_{\Gamma}
& =
( 1 - t_{\Gamma} ) \;
\Big( I_{\Gamma}
 -  \sum_{i=1}^3 I_{\Gamma/\gamma_i} \,  t_{\gamma_i} I_{\gamma_i}
 \Big)
\label{sunriseforest}
\nn\\
& =
I_\Gamma - t_{\Gamma} I_{\Gamma}
-\sum_{i=1}^3
I_{\Gamma/\gamma_i} \, t_{\gamma_i}  I_{\gamma_i}
+\sum_{i=1}^3 t_{\Gamma} \,
\big( I_{\Gamma/\gamma_i} \, t_{\gamma_i} I_{\gamma_i} \big)
\, ,
\end{align}
where the implementation of  the substitution operator $S_{\gamma _i}$ for the subgraphs $\gamma_i$
(discussed in Subsections~\ref{sec:subgraphs} and  \ref{sec:exsun})
is self-understood.

To conclude, we come back to the general case and
introduce the subtraction operator $\bar{R}_{\Gamma}$ for the
subdivergences of the graph $\Gamma$ by
\begin{align}
R_{\Gamma} I_{\Gamma} \equiv
(1 - t_{\Gamma}) \bar{R}_{\Gamma} I_{\Gamma}
\, .
\end{align}
According to \eqnref{forestformula}, the \emph{forest formula for} $\bar{R}_{\Gamma}$ then reads
\begin{align}
\label{forestformulaRB}
\fbox{\mbox{$ \
\bar{R} _\Gamma  I_\Gamma   =  \sum_{\alpha} \Big( \prod _{\gamma \in {\cal F}_{\alpha}} (- t_{\gamma} ) \Big) I_\Gamma
\ $}}
\, ,
\end{align}
where $\alpha \in \{ 0,1, \dots \}$ labels all forests ${\cal F}_{\alpha}$ which contain only non-overlapping subgraphs.

\subsection{Recursion relation for Bogoliubov's operator}\label{sec:recursion}

It can be shown~\cite{Zimmermann:1969} that the operator $\bar{R}_{\Gamma}$ satisfies the \emph{BPHZ recursion relation}
\begin{align}
\label{BPHZrecursionrel}
\fbox{\mbox{$ \
\bar{R} _\Gamma  I_\Gamma   =  \sum_{\Psi} \Big( \prod _{\gamma \in {\Psi}} (- t_{\gamma} ) \bar{R} _{\gamma} \Big) I_\Gamma
\ $}}
\, ,
\end{align}
or, equivalently
\begin{align}
\label{BPHZrecursionrelation}
\bar{R} _\Gamma  I_\Gamma  &=
\sum_{\Psi} I_{\Gamma /\Psi} \prod _{\gamma \in {\Psi}} (- t_{\gamma} ) \bar{R} _{\gamma}  I_\gamma
\, ,
\end{align}
where
$\Psi$ labels all possible sets of the form
\[
\Psi = \{ \gamma \, / \, \gamma = \mbox{disjoint renormalization part of $\Gamma$} \}
\, ,
\]
including the case $\Psi = \{ \emptyset \}$.
Before explaining the passage from (\ref{BPHZrecursionrel}) to (\ref{BPHZrecursionrelation}), we
emphasize two points.
The recursion relation  determines the operator $\bar{R} _\Gamma $ recursively in terms of the operators
 $\bar{R} _\gamma $ corresponding to lower order graphs $\gamma$.
The recursion relation only involves \emph{disjoint}
 subgraphs while the forest formula
refers to non-overlapping subgraphs, i.e. to both disjoint and nested subgraphs.

Let us now explain the expression on the right hand side of \eqnref{BPHZrecursionrelation}.
For a set $\{\gamma_1,\ldots,\gamma_n\}$ of mutually disjoint renormalization parts of the graph $\Gamma$, one defines the \emph{reduced diagram}
${\Gamma/\{\gamma_1,\ldots,\gamma_n\}}$  by contracting each $\gamma_i$ to a point.
For disjoint subgraphs $\gamma_1, \gamma_2$ of $\Gamma$,
the order of $t_{\gamma_1}$ and $t_{\gamma_2}$ does not matter and
we have
\begin{align}
 t_{\gamma_1} t_{\gamma_2}  I_\Gamma =  I_{\Gamma / \{ \gamma_1 , \gamma_2 \}} (t_{\gamma_1} I_{\gamma_1}) (t_{\gamma_2} I_{\gamma_2})
\, ,
\end{align}
and
\begin{align}
\Big( ( t_{\gamma_1}  \bar{R} _{\gamma_1} )  ( t_{\gamma_2}  \bar{R} _{\gamma_2} ) \Big)  I_\Gamma =
 I_{\Gamma / \{ \gamma_1 , \gamma_2 \}} (t_{\gamma_1} \bar{R} _{\gamma_1}  I_{\gamma_1}) (t_{\gamma_2} \bar{R} _{\gamma_2}
 I_{\gamma_2})
\, ,
\end{align}
which explains the equivalence between Equations \eqref{BPHZrecursionrel} and \eqref{BPHZrecursionrelation}.
By contrast, for nested subgraphs $\gamma_1 \subset \gamma_2$, we have to apply $t_{\gamma_1}$ first and \eqnref{fac} yields
\begin{align}
t_{\gamma_2} t_{\gamma_1}  I_\Gamma =  I_{\Gamma / \gamma_2 } \,  t_{\gamma_2} \Big( I_{\gamma_2 / \gamma_1} (t_{\gamma_1} I_{\gamma_1}) \Big)
\, .
\end{align}

The fact that the forest formula (\ref{forestformulaRB}) solves the recursion relation (\ref{BPHZrecursionrel})
can easily be checked for the example (\ref{exoverlap}): in this case,
$\Psi = \{ \emptyset \}$ or $ \Psi = \{ \gamma_1 \}$ or  $ \Psi = \{ \gamma_2 \}$, hence  (\ref{BPHZrecursionrelation})
states that
\begin{align}
\bar{R} _{\Gamma} I_{\Gamma}
= I_{\Gamma} -  I_{\Gamma / \gamma_1} t_{\gamma_1} ( \bar{R} _{\gamma_1} I_{\gamma_1})
- I_{\Gamma / \gamma_2} t_{\gamma_2} ( \bar{R} _{\gamma_2} I_{\gamma_2})
\, .
\end{align}
Since $\gamma_1$ does not contain any subdivergences, the action of the operator $\bar{R} _{\gamma_1} $
(which eliminates all subdivergences from the graph ${\gamma_1}$) on $ I_{\gamma_1}$ is trivial, i.e.  $\bar{R} _{\gamma_1} I_{\gamma_1} =  I_{\gamma_1} $
(and similarly $\bar{R} _{\gamma_2} I_{\gamma_2} =  I_{\gamma_2} $),
hence
\begin{align}
\bar{R} _{\Gamma} I_{\Gamma}
= I_{\Gamma} -  I_{\Gamma / \gamma_1} \, t_{\gamma_1} I_{\gamma_1}
- I_{\Gamma / \gamma_2} \, t_{\gamma_2}  I_{\gamma_2}
= ( 1 -  t_{\gamma_1} -  t_{\gamma_2} ) I_{\Gamma}
\, ,
\end{align}
so that
\begin{align}
R_{\Gamma} I_{\Gamma} \equiv  ( 1 -  t_{\Gamma} ) \bar{R} _{\Gamma} I_{\Gamma}
= ( 1 -  t_{\Gamma} ) ( 1 -  t_{\gamma_1} -  t_{\gamma_2} ) I_{\Gamma}
\, ,
\end{align}
i.e. the forest formula (\ref{exoverlap}) for the example under consideration.

\subsection{Example: Renormalization of the sunrise graph}\label{sec:renormsunrise}

The sunrise graph involves two interaction vertices, i.e.
a factor ${\lambda}^2$.
The unrenormalized sunrise graph is described by the integral
\begin{align}
J_{\Gamma} (p) &\equiv
\int d^4k_1  \int d^4k_2
\,
I_{\Gamma} ( p, k_1, k_2 )
\, ,
\label{unrenormintegral}
\end{align}
where the integrand is a product of
propagators (\ref{propagator}) -- see \figref{fig:Sunrise-with-momenta} :
\begin{align}
I_{\Gamma} ( p, k_1, k_2 )
 &\equiv
\tilde{\Delta} \big( \frac{p}{3}+k_1 \big)
\, \tilde{\Delta} \big( \frac{p}{3}+k_2-k_1  \big)
\, \tilde{\Delta}  \big( \frac{p}{3}-k_2  \big)
\,.
\label{prodprop}
\end{align}
(In expression (\ref{unrenormintegral}), we
suppressed the numerical prefactor involving $\lambda^2$.)
The \emph{renormalized sunrise graph} is now given by
\begin{align}
\label{renormintegral}
J ^{{\rm finite}} _{\Gamma} (p)
 &=
\int d^4k_1  \int d^4k_2
\, \big( R_{\Gamma} I_{\Gamma } \big) ( p, k_1, k_2 )
\, , \quad\nn\\
\textrm{with} \quad
R_{\Gamma} I_{\Gamma }
&\equiv  (1-t_{\Gamma} ) \, \bar R _{\Gamma} I_{\Gamma}
= (1-t_p^2) \, \bar R _{\Gamma} I_{\Gamma}
\, ,
\end{align}
where $\bar R _{\Gamma} I_{\Gamma}$ is determined by the forest formula for the sunrise graph, i.e. \eqnref{sunriseforest}:
\begin{align}
\nonumber
\big( \bar R _{\Gamma} I_{\Gamma} \big) ( p, k_1, k_2 )
&=
\tilde{\Delta} \big( \frac{p}{3}+k_1 \big)
\, \tilde{\Delta} \big( \frac{p}{3} + \!k_2 \! - \!k_1 \big)
\, \tilde{\Delta} \big( \frac{p}{3} -k_2 \big)
 -
 \tilde{\Delta} \big( \frac{p}{3}-k_2 \big)
\, \left[ \tilde{\Delta} \big( k_1-\frac{k_2}{2} \big) \right]^2
\\
&\quad
 -
 \tilde{\Delta} \big( \frac{p}{3} + \!k_2 \! - \! k_1 \big)
\, \left[ \tilde{\Delta} \big(  \frac{k_1}{2}  +  \frac{k_2}{2} \big) \right]^2
 -
 \tilde{\Delta} ( \frac{p}{3}+k_1 )
\, \left[ \tilde{\Delta} ( k_2\!-\!\frac{k_1}{2} ) \right]^2 .
\end{align}
To evaluate the  integral   $\int d^4k_1  \int d^4k_2
\, \big( \bar R _{\Gamma} I_{\Gamma} \big) ( p, k_1, k_2 ) $, one substitutes Schwinger's parametric representation for the propagators,
\begin{align}
\tilde{\Delta} ( q_i ) \equiv
\frac{1}{q_i^2+ m^2 }= \int\limits_0^\infty d\alpha_i \, \re ^{-\alpha_i (q_i^2+ m^2 )}
\, ,
\end{align}
so that the integration over $\vec k \equiv (k_1 , k_2 ) \in \R^8$ yields integrals of Gaussian type,
\[
\int d^8k \, \re ^{- \langle \vec k , A \vec k \, \rangle - \langle \vec b ,  \vec k \, \rangle }
=
\re ^{ \langle \vec b , A^{-1} \vec b \, \rangle} \, \left( \frac{\pi}{\sqrt{{\rm det} \, A}} \right) ^4
\, .
\]
Thus, one finds that
\begin{align}
J ^{{\rm finite}} _{\Gamma} (p)
& = \pi ^4
\int\limits_0^\infty d\alpha_1 \int\limits_0^\infty d\alpha_2 \int\limits_0^\infty d\alpha_3
\,
\re^{-(\alpha_1 + \alpha_2 +\alpha_3) m^2 }
(1-t_p^2) \,
\Bigg\{
\frac{ \re^{ - \beta \, p^2} }{(\alpha_1 \alpha _2 + \alpha_2 \alpha _3 + \alpha_1 \alpha _3 )^2}
\nn \\
& \hspace*{3.9cm}
- \frac{1 }{ \alpha_1^2 (\alpha_2 + \alpha_3) ^2 }
- \frac{1 }{ \alpha_2^2 (\alpha_1 + \alpha_3) ^2 }
- \frac{1 }{ \alpha_3^2 (\alpha_1 + \alpha_2) ^2 }  \Bigg\}
\, ,
\label{triple}
\end{align}
where
$\beta \equiv \frac{ \alpha_1 \alpha _2  \alpha _3  }{\alpha_1 \alpha _2 + \alpha_2 \alpha _3 + \alpha_1 \alpha _3 }$.
Finally, from
\[
- \inv2 \, p^{\mu} p^{\nu} \, \left.\frac{\partial ^2 \re^{-  \beta \, p^2}}{\partial p^{\mu} \partial p^{\nu}} \right| _{p=0}
= \, \beta \, p^2
\, ,
\]
we conclude that the \emph{renormalized sunrise graph} is given by
\begin{align}
\label{renormsunrisegraph}
J ^{{\rm finite}} _{\Gamma} (p)
 & =
\pi^4
\, \int\limits_0^\infty d\alpha_1 \int\limits_0^\infty d\alpha_2 \int\limits_0^\infty d\alpha_3
\, \frac{\re^{-(\alpha_1 + \alpha_2 +\alpha_3) m^2 }}{  (\alpha_1 \alpha _2 + \alpha_2 \alpha _3 + \alpha_1 \alpha _3 )^2  }
\left\{
\re^{-  \beta \, p^2} - (1 - \beta p^2)
\right\}
\, .
\end{align}

\subsection{Relationship to additive (and multiplicative) renormalization}\label{sec:reladd}

Let us investigate how the subtractions performed for the sunrise graph $\Gamma$ affect
the initial Lagrangian of the $\phi^4$-theory~\cite{fredenhagen}.
For the quadratically divergent graph $\Gamma$ we expand the integrand $I \equiv I_{\Gamma}$
in a Taylor series with respect to the external momentum $p$ around $p=0$:
\begin{align}
I(p, k_1, k_2) & = \big( t_p^2 I \big) (p, k_1, k_2) + R    (p, k_1, k_2)
\, .
\end{align}
Here, $t_p^2$ denotes as before the Taylor series expansion operator up to order $2$ in $p$,
and the remainder term $R$ decays strong enough with respect to $k_1$ and $k_2$ that it yields
a finite contribution to the Feynman integral
$\int d^4k_1 \int d^4k_1 \, I    (p, k_1, k_2)$.
By explicitly spelling out the terms of order $0, 1$ and $2$ in $p^{\mu}$
which make up the expansion
\begin{align}
\big( t_p^2 I \big) (p, k_1, k_2)
= t_p^2 \Big( \tilde{\Delta} \big( \frac{p}{3}+k_1 \big)
\, \tilde{\Delta} \big( \frac{p}{3} + \!k_2 \! - \!k_1 \big)
\, \tilde{\Delta} \big( \frac{p}{3} -k_2 \big)
 \Big)
\, ,
\end{align}
one notes that these terms  yield a quadratic, linear and logarithmic divergence in the Feynman integral, respectively.
To handle these divergences, we introduce a cut-off $\Lambda >0$ in momentum space
and a smooth test function $( k_1, k_2) \mapsto f (k_1, k_2)$
of compact support (i.e. $f \in {\cal D} (\R ^8)$) satisfying $f(0,0 )=1$. To regularize the integrand $I$, we now smear out with
the function $f_{\Lambda}$ defined by
\begin{align}
f_{\Lambda} ( k_1, k_2) \equiv f \Big( \frac{k_1}{\Lambda}, \frac{k_2}{\Lambda} \Big)
\, ,
\end{align}
which tends to $1$ as the cut-off $\Lambda$ goes to infinity,
i.e. we consider the \emph{regularized integral}
\begin{align}
J_{\Lambda} (p) &\equiv \int d^4k_1 \int d^4k_1 \,  f_{\Lambda} ( k_1, k_2) \,  I    (p, k_1, k_2)
\nn
\\
&= a(\Lambda) +  b_{\mu} (\Lambda)\,  p^{\mu} +  c_{\mu \nu} (\Lambda) \, p^{\mu} p^{\nu} + R (p, \Lambda )
\, .
\label{regcutoff}
\end{align}
For $\Lambda \to \infty$, we have $a(\Lambda) \propto \Lambda^2, \,   b_{\mu} (\Lambda) \propto \Lambda$ and $c_{\mu \nu} (\Lambda)
\propto \ln \Lambda$, while $R (p, \Lambda )$ yields a regular function $R(p)$.

The freedom in the choice of the test function
$f$ can be exploited to achieve $b_{\mu}=0$ and $c_{\mu \nu} (\Lambda) = c (\Lambda) \delta_{\mu \nu}$.
The subtraction of the divergent part $ t_p^2 I$ from the integrand $I$, which was considered in the BPHZ approach, therefore amounts to adding
to the momentum space action the counterterm
\begin{align}
\inv2 \int
\frac{d^4p}{(2\pi)^4}
 \, \big( a + c \, p^2 \big)
 \tilde{\phi} ^2
=
\inv2
  \int d^4x \, \Big(  c
  \, \pa^{\mu} \phi \pa_{\mu} \phi +a \phi^2 \Big)
\, ,
\end{align}
i.e. we have a local counterterm for the (quadratic part of the) Lagrangian
which has the same form as the initial action (``additive renormalization'').

Up to the considered order of perturbation theory, the  (quadratic part of the)
redefined Lagrangian thus has the form
 \begin{align}
 {\cal L} _{\rm ren} =
\inv2 \, \big( 1 + c(\Lambda) \big) \pa^{\mu} \phi \pa_{\mu} \phi +
\inv2 \, \big( m^2 + a(\Lambda) \big) \phi^2
\, .
\end{align}
With
\begin{align}
Z(\Lambda) \equiv  1 + c(\Lambda)
\, , \quad
\phi_{\rm ren} \equiv \sqrt{Z(\Lambda)} \, \phi
\, , \quad
m_{\rm ren}^2 \equiv \big( m^2 + a(\Lambda) \big) \, Z(\Lambda) ^{-1}
\, ,
\end{align}
we have  (``multiplicative renormalization'')
\begin{align}
 {\cal L} _{\rm ren} =
\inv2 \,  \pa^{\mu} \phi_{\rm ren} \pa_{\mu} \phi_{\rm ren} +
\inv2 \,  m_{\rm ren} ^2  \phi_{\rm ren} ^2
\, .
\end{align}

After the renormalizations have been performed, it is still possible to add finite ($\Lambda$-independent) terms
to $a(\Lambda )$ and  $c(\Lambda )$: these finite renormalizations have been encountered in Subsection~\ref{subtractoperator}
as the ambiguity of extracting the finite part $J_{\Gamma}^{{\rm finite}}(p)$ from the quadratically divergent
integral $J_{\Gamma} (p)$.

As we noted in Subsection~\ref{subtractoperator},
the subtraction (\ref{sublog}) for the logarithmically divergent fish diagram
(\ref{4ptfct}) leads to
the addition of a local counterterm of the form $d(\Lambda ) \phi^4$ to the Lagrangian,
i.e. a contribution of the same form as the interaction term $\frac{\lambda}{4!} \, \phi^4$
which is present in the initial Lagrangian.

For the comparison with (and relationship to)
 other approaches to renormalization, we refer to the work~\cite{Sampaio:2002ii,Falk:2009ug}.

\subsection{Locality of subtractions and renormalizability}\label{sec:locality}

The application of the forest formula to each graph $\Gamma$ allows to render all Feynman integrals
convergent. The second step of the BPHZ renormalization procedure
consists of showing that the performed subtractions
are equivalent to the addition of local counterterms to the Lagrangian. The consideration of the
forest formula is not judicious for establishing this equivalence since it does not explicitly refer to
the different orders of perturbation
theory~\cite{muta}.
As discussed in Subsection~\ref{sec:recursion},
the forest formula solves
\emph{the BPHZ recursion relation,}
and the latter
\emph{relates the subtractions performed at different orders of perturbation theory}
so that this relation can be used
to establish locality of the renormalization process.  In fact, whatever the graph or subgraph $\Gamma$ to which the recursion relation
(\ref{BPHZrecursionrel})
or (\ref{BPHZrecursionrelation})
is applied to, it relates $\bar R _{\Gamma}$ to $\bar R_{\gamma}$ for subgraphs $\gamma$
and thereby different orders of perturbation theory.
Accordingly, if one can show that the field theoretic model under consideration is renormalizable at one-loop order
(i.e. the BPHZ subtraction for divergent one-loop graphs can be implemented by the addition of a local counterterm),
then the locality of the counterterms at two-loop order follows from the recursion relation
for the two-loop subgraphs $\Gamma(2)$ of a given graph $\Gamma$:
\begin{align}
\label{BPHZlocal}
\bar{R} _{\Gamma (2)}  I_{\Gamma (2)} & =  \sum_{\Psi (1)}
I_{\Gamma (2) / \Psi (1)}   \prod _{\gamma (1) \in {\Psi (1)}} \big(- t_{\gamma (1) } \big) \, I_{\gamma (1)}
\, ,
\end{align}
where $\Psi (1)$ labels the sets of disjoint one-loop renormalization parts
and where we used the fact that $\bar{R} _{\gamma (1) }  I_{\gamma (1)} =  I_{\gamma (1)} $.
By virtue of  (\ref{BPHZlocal}) all one-loop subdivergences are eliminated in the two-loop subgraphs $\Gamma (2)$.
By proceeding along the same lines for higher order subgraphs of $\Gamma$ and applying the operator $(1 - t_{\Gamma})$
 in the end to subtract a possible overall divergence of $\Gamma$, one establishes the locality of counterterms
 and thereby shows the equivalence of the BPHZ approach to renormalization to the
 method based on adding counterterms to the Lagrangian.

\subsection{Case of massless fields (\texorpdfstring{$s$}{s}-trick)}\label{sec:massless}

In view of the treatment of gauge fields we comment on the $\phi^4$-theory
for a massless field. In this case, potential IR problems appear. For instance,
 the integral (\ref{4ptfct}) corresponding to the fish diagram then reduces to
 $\intk \,  \frac{1}{k^2 (p+k)^2 }$:
this integral admits an IR-divergence
for $p=0$ which is precisely the value for which the BPHZ subtraction  is performed.
Thus, in the case of massless fields, the standard BPHZ subtraction scheme considered for
discarding UV-divergences may introduce some artificial
IR-singularities~\cite{Lowenstein:1974uk, Lowenstein:1974qt,Lowenstein:1975ug, Lowenstein:1975pd,  Lowenstein:1975}.
The remedy consists of making the subtraction either for a non-zero value of
$p$ or for a non-zero value of the mass.
Following Lowenstein et al.~\cite{Lowenstein:1974uk, Lowenstein:1974qt,Lowenstein:1975ug, Lowenstein:1975pd,  Lowenstein:1975}
one generally adopts the latter procedure for its simplicity.
More precisely, one introduces a mass factor
$M_s \equiv (1-s )M$, which
involves an auxiliary mass $M$ and a real auxiliary variable $s$
(the ``softness parameter"),
in the effective Lagrangian and thereby
in all internal propagators:
\begin{equation}
 \frac{1}{k^2} \; \leadsto \;
\frac{1}{k^2 + M_s^2} \, ,
\qquad {\rm with} \ \;
M_s = (1-s )M
\, .
\end{equation}
Then,
one applies the Taylor expansion operator $t_{p,s}^{\delta (\Gamma)}$ around both $p=0$ and $s=0$,
while setting $s=1$ at the end of the calculation
(the whole procedure being sometimes referred to as \emph{$s$-trick}).
The possible IR-divergences due to the vanishing mass  of the physical fields (external zero mass lines of Feynman graphs)
have to be dealt with appropriately.

Thus, for a logarithmically divergent
integral like $J_{\Gamma} (p)
= \intk \,  \frac{1}{k^2 (p+k)^2 }
 \equiv
 \intk \,   I_{\Gamma} (p,k)$,
the BPHZL subtraction amounts to considering the following expression (according to  Eqns. (\ref{standardBPHZ}),(\ref{eq:subtact-op-onep}) and
the previous arguments)
\begin{align}
J_{\Gamma}^{{\rm finite}} (p)
& \equiv
\lim_{s \to 1}
\intk \, \left[ 1 - t_{p,s}^{0}  \right] \,  I_{\Gamma} (p,k,s)
  =
\lim_{s \to 1}
  \intk \, \left[ I_{\Gamma} (p,k,s) -  I_{\Gamma} (0,k,0) \right]
\nn \\
 &= \intk \, \left(
 \frac{1}{k^2 (p+k)^2 }
  -  \frac{1}{(k^2+M^2)^2}  \right)
 \,.
\label{BPHZL}
\end{align}
Accordingly, the subtraction term involves a mass so that it is not IR-divergent despite the fact
that it involves a vanishing external momentum.

After substituting Schwinger's parametrization of the propagators into the integral (\ref{BPHZL})
and performing the resulting Gaussian integrals over the four vector $k$ (see for instance
appendix of reference~\cite{Blaschke:2008b}), one
obtains the expression
\begin{align}
\frac{1}{\pi^2} \, J_{\Gamma}^{{\rm finite}} (p)
& =
\int_0^{\infty} \frac{d\lambda}{\lambda} \,
\bigg\{
\int_0^1 \, d\xi \,
{\re} ^{ - \lambda  \xi (1- \xi) p^2  }
- {\re}^{- \lambda M^2 }
 \bigg\}
 \,, \label{eq:sample-integral1}
\end{align}
The original UV-divergence of $\Gamma$ now manifests itself by a problematic behavior of the two $\lambda$-integrals for small values
of the parameter $\lambda$. These integrals can be regularized~\cite{das} by multiplying the integrands by an exponential cut-off factor
${\re}^{-\vare/\l}$ and considering the limit $\vare \to 0$ for the resulting integral.
Proceeding along these lines one finds
\begin{align}
\frac{1}{\pi^2} \, J_{\Gamma}^{{\rm finite}} (p)
& =\ln\left(\frac{M^2}{p^2}\right)+2
\,, \label{eq:sample-integral}
\end{align}
i.e. a result which has the same form as the one obtained by expanding the BPHZ-renormalized fish diagram of the massive $\phi^4$-theory~\cite{Piguet:1995}
for small values of the mass $m$:
\begin{align}
J_{\Gamma} ^{{\rm finite}} (p ;m)
& \propto \sqrt{\frac{p^2 +4 m^2}{p^2}} \, \ln {\left[\frac{\sqrt{p^2+4 m^2}-\sqrt{p^2}}{\sqrt{p^2+4 m^2}+\sqrt{p^2}}\right]} + 2
\approx \ln \left(\frac{m^2}{p^2}\right)+2
\,.
\end{align}
The  $M$-dependent one-loop result (\ref{eq:sample-integral}) for the four-point function
is absorbed by choosing an appropriate $M$-dependent coefficient $d$
(i.e. counterterm $d \, \phi^4$) which is adjusted in such a way that the Green functions satisfy \emph{normalization conditions that do not depend on the auxiliary mass $M$}~\cite{Lowenstein:1974qt}.
Thereby, the final theory does not depend on the auxiliary mass $M$ and describes
a well defined massless quantum field theory.
For further details, e.g. a discussion of the BPHZL-renormalization of the sunrise graph within the massless $\phi^4$-theory,
we refer to the work~\cite{Lowenstein:1974qt} (in particular its conclusion).

\subsection{Assessment}\label{sec:assess}

Let us summarize once more the salient features of the BPHZ scheme~\cite{Collins:1986}.
Following Zimmermann~\cite{ZimmermannBrandeis}, the subtractions providing convergent integrals
are directly applied to the integrands, hence no regulator has to be considered.
Accordingly, this procedure exhibits the fact that the properties of the resulting quantum theory
do not depend on a regulator or on the way it is introduced.
Moreover, general mathematical theorems
based on simple properties of Taylor series
ensure that finite integrals can be constructed from divergent
ones without explicitly investigating each Feynman diagram and its divergences. The BPHZ scheme
is quite useful for discussing different issues of quantum theory like the operator product expansion~\cite{ZimmermannBrandeis,ZimmermannAnnals}.
Following Lowenstein et al.~\cite{Lowenstein:1974uk, Lowenstein:1974qt,Lowenstein:1975ug, Lowenstein:1975pd,  Lowenstein:1975}
the BPHZ approach can be adapted to tackle
massless fields though the subtractions are somewhat more complex in this case,
in particular in the presence of gauge symmetries, e.g. see Ref.~\cite{Grassi:1995wr}.

\section{Non-commutative \texorpdfstring{$\phi ^4$}{phi4}-theory}\label{nctheory}

We are now ready to turn to the {\nc} case and, after discussing general properties of the modified
BPHZ approach introduced in reference~\cite{Blaschke:2012ex}, we study
its application to the sunrise graph of {\nc} $\ph^4$-theory.

\subsection{UV/IR mixing problem (Infrared singularities)}\label{IRsingularities}

Before considering
 the BPHZ approach to the UV/IR mixing problem,
it is useful to recall briefly the origin and nature of this problem~\cite{Minwalla:1999px}.

In non-commutative field theories,
the star product leads to the presence of phase factors in various Feynman graphs.
In particular, the non-planar one-loop $2$-point function (non-planar tadpole graph) which is
given, up to a numerical factor, by
\begin{align}
\Pi_{\text{n-pl}}(\tilde p)
= \int\limits_{\mathbb{R} ^4} d^4k \,
\frac{\cos(k\p)}{k^2+m^2}
\, ,
\label{eq:loopint-1}
\end{align}
involves a phase factor of the form
$\cos (k \tilde p)$ where $p$ denotes the external momentum.
For large values of $k$, the rapid oscillations of the phase factor
have a damping effect upon integration; thus, for any $\tilde p \neq 0$,  the function (\ref{eq:loopint-1}) is finite by contrast to the
corresponding planar diagram
which does not contain a phase factor,
\begin{align}
\Pi_{\text{pl}}
= \int\limits_{\mathbb{R} ^4} d^4k \,
\frac{1}{k^2+m^2}
\, ,
\label{planaroneloop}
\end{align}
and which is quadratically UV-divergent.
Accordingly, the phase factor can be viewed as a regularization brought about
the non-commutativity of space-time~\cite{Minwalla:1999px},
i.e. an idea which is reminiscent of the historical arguments
which led Heisenberg and Snyder to consider non-commutative
space-time to overcome the problem of UV-divergences in quantum field theory~\cite{Heisenberg:1930,Pauli:1946,Snyder:1946,Snyder:1947,Yang:1947}.
However, the UV-divergent diagram (\ref{planaroneloop}) is still present in the theory
and in addition the integral (\ref{eq:loopint-1}) is singular for small values
of the external momentum (IR-divergent)~\cite{Minwalla:1999px,Blaschke:2008a},
the leading singularity being $1/{\tilde p} ^{\, 2}$:
 \begin{align}
\label{smallp}
\Pi_{\text{n-pl}}(\tilde p)
&=\frac{1}{(4\pi)^2} \bigg[ \frac{4}{\p^{\, 2}} + m^2 \ln ( m^2 \p^{\, 2} ) \bigg] +\mathcal{O}(1)
\qquad {\rm for} \ \; \tilde p \ll 1
\, .
\end{align}
To have a better understanding of this expansion, it is useful to first recall the
cut-off regularization of the integral (\ref{planaroneloop}). In this respect,
one substitutes \emph{Schwinger's parametrization} $(k^2 +m^2)^{-1} = \int_0^{\infty} d\alpha \, \re^{-\alpha (k^2 +m^2)}$
into  (\ref{planaroneloop}) and performs the Gaussian integration over $k$ in order to obtain
 \begin{align}
\label{planschwinger}
\Pi_{\text{pl}}= \int_0^{\infty} \frac{d\alpha}{\alpha^2} \, \re^{-\alpha m^2}
\, .
\end{align}
The UV-divergence of  (\ref{planaroneloop}) now amounts to the singular behaviour
of (\ref{planschwinger})  for small values of $\alpha$. The divergence structure of this expression
can be exhibited by cutting off the $\alpha$-integral at a lower limit $1 / \Lambda^2$
(where $\Lambda \gg 1$ represents a \emph{momentum cut-off),} expanding the exponential and carrying out the integration of the first few terms~\cite{das},
or equivalently by cutting off the integrand by the introduction of an exponential factor $\re^{-(\Lambda^2 \alpha)^{-1} }$:
\begin{align}
\label{planregularized}
\Pi _{ \text{pl} } & = \lim _{\Lambda \to \infty} \left( \Pi _{ \text{pl} } \right) _{\rm reg} (\Lambda ) \, , \qquad
{\rm with} \ \;
\left( \Pi _{\text{pl} } \right) _{\rm reg} (\Lambda ) \equiv
\int_0^{\infty} \frac{d\alpha}{\alpha^2} \, \re^{-\alpha m^2} \, \re^{ - (\Lambda ^2 \alpha )^{-1}}
\, .
\end{align}
The final result reflecting the quadratic UV-divergence of $\Pi _{ \text{pl} }$ (and involving a
subleading logarithmic divergence) reads
\begin{equation}
\label{planregulex}
\left(  \Pi _{ \text{pl} } \right) _{\rm reg} (\Lambda ) =
\frac{1}{(4\pi)^2} \bigg[ \Lambda^2 - m^2 \ln  \frac{ \Lambda^2}{m^2}  \bigg] +\mathcal{O}(1)
\qquad {\rm for} \ \; \Lambda \gg
1
\, .
\end{equation}

Let us now come back to the integral (\ref{eq:loopint-1}):
substitution of \emph{Schwinger's parametrization} into this integral
yields
\begin{equation}
\label{nplgraph}
\Pi_{\text{n-pl}} (\tilde p ) =
\int_0^{\infty} \frac{d\alpha}{\alpha^2} \, \re^{-\alpha m^2} \, \re^{-\frac{\tilde p^{\, 2}}{4 \alpha} }
\, ,
\end{equation}
i.e. the same integral as (\ref{planregularized}) with ${\Lambda^2}$ replaced by
$\frac{4}{\tilde p^{\, 2}}$, whence the expansion (\ref{smallp}) \cite{Minwalla:1999px,Blaschke:2008a}.

The non-planar one-loop $4$-point function
involves an integral of the form (\ref{4pfct}) which
can be discussed along the same lines:
it is a function of $\tilde p$ (and of $p$) which
is UV-finite, but involves a logarithmic IR-singularity (divergence for $\tilde p \to 0$),
the latter reflecting the logarithmic UV-divergence of the corresponding  planar diagram~\cite{das},
\[
J _{ \text{pl} } (p)  = \lim _{\Lambda \to \infty} \left( J_{ \text{pl} } \right) _{\rm reg} (p, \Lambda ) \, , \quad
{\rm with} \
\left( J _{\text{pl} } \right) _{\rm reg} (0, \Lambda ) = \pi^2
 \bigg[ \ln \frac{\Lambda^2}{m^2} -1  \bigg] +\mathcal{O}(\frac{1}{\Lambda ^2})
\quad {\rm for} \ \; \Lambda \gg 1
\, .
\]

\subsection{Modified BPHZ subtractions}
\label{sec:modifiedBPHZ}

As discussed in the previous subsection, the
\emph{non-planar} one-loop $n$-point function is
UV-finite, but suffers from an IR-singularity which is tied to the
UV-divergence of the corresponding \emph{planar} diagram.
Thus, for these non-planar graphs,
our modified BPHZ subtraction amounts to an IR-subtraction rather
than a UV-subtraction. When applying this subtraction
to an integrand $I(p, \tilde p , k)$, it is important to
consider $p$ and $\tilde p$ as independent variables and to subtract the
Taylor series expansion around $p=0$ (rather than $\tilde p = 0$ which
represents a divergence of the diagram):
\begin{equation}
\label{modifiedexpand}
I(p, \tilde p , k) = I(0, \tilde p , k) + p^{\mu} \, \frac{\pa I}{\pa p^{\mu}} (0, \tilde p , k) + \dots
\, .
\end{equation}
For the subtracted
non-planar one-loop $2$-point function, we get a vanishing result (as one also does
for the planar, UV-divergent diagram by virtue of the standard
BPHZ subtraction scheme, see \eqnref{finitenpltadpole}.
For the subtracted non-planar one-loop $4$-point
function (\ref{checkJ}), one obtains a result which is regular in $\tilde p$ \cite{Blaschke:2012ex}:
\begin{align}
\label{non_planar-result}
J^{{\rm finite}} (p )
&=- \pi ^2 \sqrt{1+\frac{4 m^2}{p^2}} \, \ln {\left[\frac{\sqrt{p^2+4 m^2}+\sqrt{p^2}}{\sqrt{p^2+4 m^2}-\sqrt{p^2}}\right]}+2\pi ^2+\cO(\p^2)
\,.
\end{align}

\subsubsection*{Finite renormalizations}

Concerning the UV-divergent planar diagrams, we note that the cut-off regularization exhibits the UV-divergence
as well as its degree (power of $\Lambda$),
the latter determining also the degree of the polynomial in $p$ which is considered for the standard BPHZ subtraction,
see equation (\ref{standardBPHZ}).
As discussed in Subsection~\ref{subtractoperator} (see \eqnref{addpolynom}),
the ambiguity involved in the standard BPHZ subtraction (corresponding to a \emph{finite} renormalization)
is a polynomial in $p$ whose order is the superficial degree of UV-divergence
of the diagram under consideration.

A non-planar diagram and the regularized version of the corresponding planar diagram have the same form up to the replacement
$\Lambda ^2 \leadsto 4/{\tilde p}^{\, 2}$ -- compare for instance equations (\ref{planregularized}) and (\ref{nplgraph}).
Hence one expects that the ambiguity involved in the modified BPHZ subtraction amounts to a polynomial in
$1/\tilde p ^{\, 2}$ whose degree is determined by the degree of the IR-singularity of the
non-planar graph.
To confirm this expectation, we consider the expansion (\ref{modifiedexpand}) for the modified BPHZ subtraction.
The ambiguity is a polynomial in $p$ (with coefficients depending on the parameter $\tilde p^{\, 2}$
which is considered as an independent variable), the degree of this polynomial coinciding
with the degree of the IR-singularity of the non-planar graph.
All coefficients of this polynomial must have the correct dimension -- see the discussions following equations
(\ref{finitenpltadpole}) and (\ref{regcutoff}). For the non-planar tadpole graph,
which has a quadratic IR-singularity, we thus get a term
$A \, \phi ^2$ (with $A$
having the dimension of
a mass squared) and a term  $(\pa^{\mu} \phi)(\pa_{\mu} \phi)$,
but there is a further possibility involving $\tilde p ^{\, 2}$. In fact, the quantities $\theta^{\mu \nu}$ parameterizing
non-commutative space have the dimension of length squared (${[\hat{x} ^{\mu} , \hat{x} ^{\nu}] } = \ri \theta^{\mu \nu} \id$)
and thereby yield an extra term as ambiguity for the subtraction, namely
$\tilde{\phi} \, ({\tilde p}^{\, 2})^{-1} \tilde{\phi}$, or in configuration space,
\begin{equation}
\label{extranonlocal}
\phi \;
\frac{{a'}^2}{\wsq} \, \phi
\, ,
\qquad {\rm with} \ \;
\wsq\equiv\widetilde{\pa}^\m\widetilde{\pa}_\m=\th^{\m\m'}\th_\m^{\ \n'}\pa_{\m'}\pa_{\n'}
\, ,
\end{equation}
 where  $a'$ represents a real dimensionless constant.
 Such a non-local term is admissible in a translation invariant scalar field theory
 on non-commutative space~\cite{Gurau:2009,Blaschke:2012ex}
 and according to the familiar lines of renormalization theory, it must be included if it is not present in
the initial Lagrangian.
Actually~\cite{Blaschke:2012ex}, it is the only non-local counterterm
which can appear in a translation invariant non-commutative scalar field model.

\subsubsection*{Renormalization of the theory}

After including the term (\ref{extranonlocal}) into the Lagrangian,
the propagator  for the $\phi^4$-theory reads
\begin{align}
G(k)= & \;
\inv{k^2+m^2+\frac{a^2}{k^2}}
\, .
\label{eq:propagator}
\end{align}
It has a ``damping'' behaviour for
vanishing momentum~\cite{Gurau:2009},
\begin{equation}
\label{damp}
 \lim\limits_{k \to 0}G(k)=0\,,
\end{equation}
which allows to overcome  potential IR-divergences in higher loop graphs.
In fact, the IR-divergence of the non-planar tadpole graph $\Pi_{\text{n-pl}}(\tilde p)$ becomes potentially problematic when this graph is inserted into a loop of another diagram (e.g. the non-planar tadpole graph itself), since the external momentum of the insertion then becomes the internal momentum $k$ over which one integrates : the divergence for $k \to 0$ then represents a potential problem for the renormalizability.
However, the damping behaviour (\ref{damp})
allows to overcome this problem~\cite{Gurau:2009,Blaschke:2008a} and indeed it has been proven to provide a renormalizable model~\cite{Gurau:2009}.

\subsection{Renormalization of the sunrise graph}\label{sec:ffphase}
We now demonstrate that our modified BPHZ scheme also works
for graphs  with overlapping divergences using the example of the sunrise graph.

The sunrise graph involves two interaction vertices of the form (\ref{F9}), i.e.
a factor $\bar{\lambda}^2$.
By expanding this factor and linearizing the squares of the involved trigonometric functions~\cite{Micu:2000}
and by using the assignments of momenta  specified in \figref{fig:Sunrise-with-momenta}
we get the result
\begin{align}
\bar \lambda^2 & =
\frac{\lambda^2}{9} \bigg[1 + 2 \cos( k_1\tilde p)
+ 2 \cos\!\left( k_1\tilde k_2 +     \tfrac{1}{3}(2k_1-k_2)\tilde p \right)
+\cos\!\left(  k_1\tilde k_2 + \tfrac{2}{3}(k_2-2k_1)\tilde p \right)\bigg]
\nn \\
&\equiv  \frac{\lambda^2}{9} \big[ 1 + f(\tilde p, k_1, k_2 ) \big]
\, .
\end{align}
Thus, the  unrenormalized sunrise graph is described by the integral
\begin{align}
J_{\Gamma} (p) &\equiv
\int d^4k_1  \int d^4k_2
\, [ 1+  f(\tilde p, k_1, k_2 ) ]
\,
I_{\Gamma} ( p, k_1, k_2 )
\, ,
\label{unrenormintegral2}
\end{align}
where the
function $I_{\Gamma}$
 is the  product of propagators
(\ref{prodprop}), and where we suppressed the numerical prefactor involving $\lambda^2$.
According to our modified BPHZ subtraction scheme, the phase factor
(which depends on $\tilde p$) is not affected by the
subtraction, i.e.  the \emph{renormalized sunrise graph} is given by
\begin{align}
\label{renormintegralnc}
J ^{{\rm finite}} _{\Gamma} (p)
 &=
\int d^4k_1  \int d^4k_2
\, [ 1+  f(\tilde p, k_1, k_2 ) ]
\, \left( R_{\Gamma} I_{\Gamma} \right) ( p, k_1, k_2 )
\, ,
\end{align}
where the renormalized integrand is determined by the forest formula,
see Eqns. \eqref{renormintegral}-\eqref{renormsunrisegraph}.

Since the modulus of the phase factor $ [ 1+  f(\tilde p, k_1, k_2 ) ]$ in integral
(\ref{unrenormintegral2}) is bounded above by $6$,
the modulus of the
integral (\ref{renormintegralnc}) of
 {\nc} $\phi^4$-theory is bounded by the integral for the sunrise graph
in  commutative  $\phi^4$-theory:
\begin{align}
\left|
J ^{{\rm finite}} _{\Gamma} (p)
\right|
& \leq 6 \int d^4k_1  \int d^4k_2
\, \left|  \left( R_{\Gamma} I_{\Gamma} \right) ( p, k_1, k_2 ) \right|
\, .
\end{align}
By virtue of equation (\ref{renormsunrisegraph}),
we therefore have
\begin{align}
\left|
J ^{{\rm finite}} _{\Gamma} (p)
 \right| & \leq
\, 6 \pi^4
 \int\limits_0^\infty\! d\alpha_1 \int\limits_0^\infty\! d\alpha_2 \int\limits_0^\infty\! d\alpha_3
\, \frac{\re^{-(\alpha_1 + \alpha_2 +\alpha_3) m^2 }}{  (\alpha_1 \alpha _2 + \alpha_2 \alpha _3 + \alpha_1 \alpha _3 )^2  }
\left| \left\{
\re^{-  \beta \, p^2} - (1 - \beta p^2)
\right\} \right|
.
\end{align}

\section{Non-commutative gauge field theories}\label{ncgt}

One of the motivations for generalizing the BPHZ approach to the non-commutative setting is to develop a tool for the renormalization of {\nc} gauge theories
since the usual approaches
such as multiscale analysis break gauge invariance, e.g. see reference~\cite{Rivasseau:2007a}
for a review.
In the following, we indicate how the modified BPHZ method applies to gauge theories
while deferring to a separate work a more complete treatment of the numerous technical details to be investigated.

The ``na\"ive'' gauge field action on {\nc} Euclidean space
is given by
\begin{align}
 S_{\textrm{YM}} [A] &= \inv4
  \intx \, F_{\m\n}\star F^{\m\n}\,,
  \quad {\rm with} \ \;
 F_{\m\n} \equiv
 \pa_\m A_\n-\pa_\n A_\m-\ig\starco{A_\m}{A_\n}
 \,. \label{eq:gaugefield-naiv}
\end{align}
This functional has
to be gauge fixed and supplemented by
an adequate  ghost contribution.
The resulting model is
 independent of the chosen gauge fixing  and
  again exhibits {\uim}, hence it is
  non-renormalizable (see e.g.~\cite{Blaschke:2010kw} and references therein).
  Thus, the action has
to be modified and, inspired by the
results achieved for
the scalar models, various approaches have been
proposed in recent years~\cite{Grosse:2007,Wallet:2007c,Blaschke:2007b,Blaschke:2008a,Vilar:2009,Blaschke:2009e,Blaschke:2010ck} -- see
also the discussion~\cite{Blaschke:2013gha} and references therein.
However,
so far none of these models could be
 proven to be renormalizable, in part due to the lack of a renormalization scheme
 which is compatible with both non-commutativity and gauge symmetry.

Let us take a closer look at the \emph{one-loop vacuum polarization}
while considering the Feynman gauge fixing. Three 
Feynman graphs contribute~\cite{Blaschke:2013gha} and
after symmetrization with respect to the indices $\mu, \nu$,
their sum reads 
as follows for the na\"ive gauge field model:
\begin{align}
 \Pi_{\m\n}(p) &= g^2
  \!\int\! \frac{d^4k}{(2\pi)^4} \frac{1\!-\!\cos(k\p)}{k^2(k+p)^2}
   \Big[8k_\m k_\n-2p_\m p_\n + 4 (p_\m k_\n
+  p_\n k_\m )-\d_{\m\n} ( p^2 +4k^2 +10 pk ) \Big]  .
\label{vacpol3}
 \end{align}
 By virtue of the change of variables $k \leadsto -(k+p)$ and the fact that
 $p\p=0$ (which follows from the antisymmetry of $\th^{\m\n}$), we get the relation
 \begin{align}
 \int\! \frac{d^4k}{(2\pi)^4} \frac{1\!-\!\cos(k\p)}{k^2(k+p)^2}
 \, (p_\m k_\n +  p_\n k_\m ) =
 -  \int\! \frac{d^4k}{(2\pi)^4} \frac{1\!-\!\cos(k\p)}{k^2(k+p)^2}
\, p_\m  p_\n
 \, ,
 \end{align}
 which implies that expression (\ref{vacpol3}) takes the simpler form
\begin{align}
 \Pi_{\m\n}(p)
  &=2g^2\!\int\!\frac{d^4k}{(2\pi)^4}\frac{1\!-\!\cos(k\p)}{k^2(k+p)^2}\Big[4k_\m k_\n-3p_\m p_\n+2\d_{\m\n} (p^2-k^2) \Big]
  \nn\\ &
 =:\intk\, I_{\G \mu \nu}(p,\p,k)
 \,. \label{vacuumpolar}
\end{align}
The phase independent part (i.e. the one
which does not involve
the cosine) is superficially quadratically
UV-divergent
by power counting, however it is well known that gauge symmetry (i.e. the Ward identity $p^\m \Pi_{\m\n}=0$) reduces this degree of divergence to
a logarithmic one.
On the other hand, the phase dependent contribution is UV-finite due to the regularizing effect of the
cosine, but it
develops a quadratic IR-singularity
for $\p\to 0$:
we have
\begin{align}
 \Pi_{\m\n}
 &=\frac{2g^2}{\pi^2}\frac{\p_\m \p_\n}{(\p^{\, 2})^2}
 \qquad
 {\rm for} \ \; \tilde p^{\, 2} \ll 1
 \, . \label{eq:gauge-IR-div}
\end{align}
This IR-divergence remains a quadratic one since it is compatible with the Ward identity $p^\m \Pi_{\m\n}=0$ following from the gauge symmetry
due to the fact that $p\p=0$.

As discussed in \secref{sec:massless}, massless theories require additional regularization in the infrared regime.
However, such a regularization is potentially
problematic for gauge models
since a regulator mass generically violates gauge invariance\footnote{Actually, the $s$-trick has recently also been implemented via a BRST-doublet, see~\cite{Quadri:2003pq}.} -- see the discussion in \appref{sec:appendix} and
references~\cite{Lowenstein:1975pd,Grassi:1995wr}.
In the commutative case, this issue is usually addressed
by using dimensional regularization which however is not appropriate
in the {\nc} setting, in particular due to the {\uim}.
 Furthermore and more problematically,
the IR-divergences of the type (\ref{eq:gauge-IR-div}) arise from the UV-divergences (i.e. the infamous
 {\uim} problem) and are at the origin of the
non-renormalizability of the na\"ive gauge field model determined by the action (\ref{eq:gaugefield-naiv}). Therefore, we will consider a gauge field model
with additional terms in the action~\cite{Blaschke:2009e} which 
provide a damping in the infrared regime
for the gauge field propagator similar to the one for the scalar $1/p^2$ model of Gurau \etal~\cite{Gurau:2009}.
Thus, the one-loop vacuum polarization in a Feynman-like gauge fixing becomes
$\Pi^{(a)}_{\m\n}(p)
\equiv  \intk\, I^{(a)}_{\G \mu \nu}(p,\p,k)$ with\footnote{For the sake of simplicity,
 we assume that the parameter $\sigma$ appearing in the  gauge field propagator of Ref.~\cite{Blaschke:2009e} vanishes,
 i.e.  in the following calculation we neglect an extra non-local counterterm for the singularity \eqref{eq:gauge-IR-div}.
Furthermore, for the present illustration we consider a Feynman-like gauge fixing where an additional damping factor is included in order to arrive at the simplest form of the gauge field propagator.
We note, however, that the full model of Ref.~\cite{Blaschke:2009e} is based on the Landau gauge fixing, or may be generalized to other gauges along the lines of the recent work~\cite{Lavrov:2013boa}.}
\begin{align}
 I^{(a)}_{\G\m\n}(p,\p,k)& \equiv \frac{2g^2}{(2\pi)^4} \,
 \frac{1-\cos(k\p) }{ k^2+\frac{a^2}{k^2}} \;
 \frac{4k_\m k_\n-3p_\m p_\n+2\d_{\m\n} (p^2-k^2)}{(k+p)^2+\frac{a^2}{(k+p)^2}}
 \,.
\end{align}
According to Eqns. (\ref{standardBPHZ}),(\ref{eq:subtact-op-onep}) and (\ref{modifiedexpand}),
we have to evaluate
\begin{align}
  &\intk \, (t_{p}^2 I^{(a)}_{\G \mu \nu} )(p,\p,k)
  =\intk \,
   \bigg(
 I^{(a)}_{\G \mu \nu} (0,\p,k)  +
  p^{\rho} \, \diff{ I^{(a)}_{\G \mu \nu} }{ p^{\rho} } (0,\p,k)
 +  \frac{p^{\rho} p^{\sigma}}{2}
\,  \ddiff{I^{(a)}_{\G \mu \nu}}{p^{\rho}}{p^{\sigma}}(0,\p,k)
\bigg)
  \,.
\end{align}
Taking into account the fact that the integral over an odd function of $k$ vanishes upon integration over symmetric intervals we find that
\begin{align}
 \label{intpun}
  \Pi_{\m\n}^{(a) \textrm{finite}}(p) &\equiv  \intk \left(1-t_{p}^2\right)
  I^{(a)}_{\G \mu \nu}
   (p,\p,k) \\
   &={2g^2}\!\int\!\frac{d^4k}{(2\pi)^4}
 \, \frac{1\!-\!\cos(k\p)}{N} \,
   \Bigg\{
   \frac{4 k_\m k_\n-2 \d_{\m\n} k^2}{N^2}
 \Bigg[p^2-\frac{a^2p^2}{\left(k^2\right)^2}+\frac{4(kp)^2}{N}  \left(\frac{3a^2}{(k^2)^2}\!-\!1\right) \!\!\Bigg]\nn\\
 &\qquad \qquad +\left[4k_\m k_\n-3p_\m p_\n+2\d_{\m\n}(p^2-k^2)\right] \Bigg[\inv{(k\!+\!p)^2+\frac{a^2}{(k+p)^2}}-\inv{N}\!\Bigg] \Bigg\}
   \,,  \nn
\end{align}
where we introduced the abbreviation
\begin{align}
 N&:=k^2+\frac{a^2}{k^2}
 \,.
\end{align}
The integral (\ref{intpun}) may eventually be carried out further by using the decomposition~\cite{Blaschke:2008b,Blaschke:2009a}
\begin{align}
\left( k^2+\frac{a^2}{k^2} \right)^{-1}&=
 \inv2\sum_{\zeta=\pm1}\frac{1}{k^2+\ri a\zeta}
 \,.
\end{align}
However, the main point is that expression (\ref{intpun}) represents a UV- and IR-finite result.

\section{Conclusion}
\label{sec:con}

By considering the sunrise graph as a prototype example, we have shown that the modified BPHZ scheme put forward in reference~\cite{Blaschke:2012ex}
works for higher loop graphs involving overlapping divergences
and that its application is unambiguous.
Furthermore, we have addressed the UV/IR mixing problem in this approach and shown that the modified BPHZ scheme yields well defined results.
According to the familiar rules of renormalization theory, this scheme implies the introduction
of a non-local term into the action.
The nature of this non-locality is precisely the one allowed (and induced) by the star product.
 The resulting Lagrangian has previously been shown to define a renormalizable theory
by application of multiscale analysis~\cite{Gurau:2009}.
The application of the modified BPHZ scheme to {\nc} gauge field theories looks promising, but
a more complete treatment  requires further investigations.

\subsection*{Acknowledgements}
D.N. Blaschke is a recipient of an APART fellowship of the Austrian Academy of Sciences, and is also grateful for the hospitality of the theory division of LANL and its partial financial support.
F. Gieres wishes the express his gratitude to S. Theisen for valuable discussions.
M. Schweda thanks C. Becchi for useful comments.

\appendix
\section{Appendix}
\label{sec:appendix}
The purpose of this appendix is to illustrate problems arising
for the one-loop polarization (\ref{vacuumpolar}) associated to the
na\"ive {\nc} gauge field model (\ref{eq:gaugefield-naiv})
when applying the \emph{$s$-trick} described in \secref{sec:massless}.
Thus, we introduce a mass term involving an auxiliary mass $M$ and an auxiliary variable $s$ in all internal propagators,
\begin{equation}
\frac{1}{k^2} \; \leadsto \; \frac{1}{k^2 + M_s^2}
\, , \qquad {\rm with} \ \; M_s \equiv (1-s) M
\, ,
\end{equation}
and apply the Taylor expansion operator $t_{p,s}^{\delta (\Gamma)}$ around both $p=0$ and $s=0$,
while setting $s=1$ at the end of the calculation so as to discard artificial IR-problems.

For the quadratically divergent integral (\ref{vacuumpolar}) we have
\begin{align}
  &\intk \, (t_{p,s}^2 I_{\G \mu \nu} )(p,\p,k,s)\nn\\
  &=\intk \,
   \bigg(
 I_{\G \mu \nu} (0,\p,k,0)  + \cdots +  \frac{p^{\rho} p^{\sigma}}{2}
  \, \ddiff{I_{\G \mu \nu}}{p^{\rho}}{p^{\sigma}}(0,\p,k,0)
   +  \frac{s^2}{2} \, \frac{\partial ^2 I_{\G \mu \nu} }{\partial s^2}
(0,\p,k,0)\bigg)
  \nn\\
  &={2g^2}\!\int\!\frac{d^4k}{(2\pi)^4} \frac{1\!-\!\cos(k\p)}{(k^2+M^2)^2}
  \bigg\{ 4k_\m k_\n-3p_\m p_\n+2\d_{\m\n} (p^2-k^2) \nn\\
  &\qquad\qquad\qquad\qquad +2
  \, \frac{2{k_\m k_\n}- \d_{\m\n} k^2 }{k^2+M^2}
\,
 \left[ 4 \,   \frac{ (pk)^2+3 s^2 M^4 }{k^2+M^2}
 + 2 s(2-s) M^2  -p^2\right]
   \bigg\}
  \,,
\end{align}
where we took into account the fact that the integral over an odd function of $k$ vanishes upon integration over symmetric intervals.
Thus, the finite part of the vacuum polarization reads
\begin{align}
  &\Pi_{\m\n}^{\textrm{finite}}(p) = \lim_{s\to 1}
  \intk\left(1-t_{p,s}^2\right)
  I_{\G \mu \nu}
   (p,\p,k,s) \nn\\
   &={2g^2} \lim_{s\to 1}
   \int\!\frac{d^4k}{(2\pi)^4} {[ 1\!-\!\cos(k\p)]}
   \bigg\{\!\!
   -2
  \, \frac{2{k_\m k_\n}- \d_{\m\n} k^2 }{(k^2+M^2)^3}
 \left[ 4 \,   \frac{ (pk)^2+3 s^2 M^4 }{k^2+M^2}
 + 2 s(2-s) M^2 -p^2\right]
    \nn\\
   &\quad\quad +
   \left[ 4k_\m k_\n-3p_\m p_\n+2\d_{\m\n}(p^2-k^2)\right]
   \left[ \inv{(k^2+M_s^2)\left((k+p)^2+M_s^2\right)}-\inv{\left(k^2+M^2\right)^2} \right] \bigg\}
   \,.
\end{align}
Evaluation of the resulting integral using Schwinger's parametrization
 leads to an expression whose planar part
is UV-finite,
and whose non-planar part involves modified Bessel functions of the second kind.
Expansion of the latter around small values of
$\p^2$ enables us to perform the final
parametric  integral (i.e. the integral
over the Schwinger parameter $\xi$ already considered in
 \eqref{eq:sample-integral1}).
 After an
 expansion around small mass $M$ we finally get
\begin{align}
  \Pi_{\m\n}^{\textrm{finite}}(p)
 &=
 \frac{g^2 M^2}{96 \pi ^2}
 \Big\{  2p^2 \p_{\mu } \p_{\nu }  \left[ \log \left(\tinv4 M^2 \p^2\right)+2 \gamma_E  -1\right]
   \nn\\
 &\quad\qquad +5 \p^2 (p_{\mu } p_{\nu }-p^2\d_{\m\n}) \left[ \log \left(\tinv4 M^2 \p^2\right)+2 \gamma_E
   -2\right]
   \nn\\
 &\quad\qquad +p^2\p^2 \delta _{\mu  \nu } \left[ 1-2 \gamma_E- \log \left(\tinv4 M^2 \p^2\right)\right] \Big\}
 \nn\\
   &\quad -\frac{g^2 p^2\p^2}{4800\pi^2}(p_\m p_\n-p^2\d_{\m\n})\left[ 45\log\left(\tinv4p^2\p^2\right)+90\g_E-163\right]
   \nn\\
   &\quad -\frac{g^2 \left(p^2\right)^2}{7200 \pi ^2} \p_{\mu } \p_{\nu } \left[ 15 \log \left(\tinv4
   p^2 \p^2\right)+30 \gamma_E -46\right]
   +\cO{\left((\p^2)^2,M^4\right)}
   \,.
\end{align}
The mass dependent parts are not transversal, but the limit $M \to 0$ exists,
\begin{align}
 \lim_{M\to0}
 \Pi_{\m\n}^{\textrm{finite}}(p)
 &=-\frac{g^2 p^2\p^2}{4800\pi^2}(p_\m p_\n-p^2\d_{\m\n})\left[ 45\log\left(\tinv4p^2\p^2\right)+90\g_E-163\right]
 \nn\\
 &\quad -\frac{g^2 \left(p^2\right)^2}{7200 \pi ^2} \p_{\mu } \p_{\nu } \left[ 15 \log \left(\tinv4
   p^2 \p^2\right)+30 \gamma_E -46 \right]
   +\cO{\left((\p^2)^2\right)}
 \,,
\end{align}
and this expression
 is indeed transversal.
These results show already at one-loop level, that the introduction of a regulator mass explicitly breaks gauge invariance and
violates
Slavnov-Taylor identities such as the
transversality of the vacuum polarization, see also~\cite{Grassi:1995wr} and references therein.
In our specific example, the limit $M\to0$ exists
and restores gauge symmetry, but this need not  be the case for other graphs.
Fortunately,
as discussed in \secref{ncgt}, we do not
have to consider an infrared regularization using an auxiliary mass  for
  the {\nc} gauge field model of Ref.~\cite{Blaschke:2009e}.



\end{document}